\documentclass[12pt,reqno]{amsart}

\usepackage{fullpage}
\usepackage{graphicx}
\usepackage{amssymb, amsmath, amsxtra}
\usepackage{enumerate}

\newtheorem{thm}{Theorem}[section]
\theoremstyle{plain}

\theoremstyle{plain}

\theoremstyle{plain}

\newtheorem{prop}{Proposition}[section]

\def\X{\boldsymbol{\rm{X}}}
\def\Y{\boldsymbol{\rm{Y}}}
\def\pr{{\rm pr}}

\def\Rnum{{\mathbb R}}
\def\Esp{{\mathcal{E}}}
\def\C{{\mathcal{C}}}
\def\G{{\mathcal{G}}}
\def\g{{\mathfrak{g}}}
\def\Gsolv{\G_{\text{solv}}}
\def\gsolv{\g_{\text{solv}}}
\def\vol{{\text{vol}}}

\def\spn{{\rm span}}

\def\parder#1{\partial_{#1}}

\def\const{\text{const.}}
\def\smallint{{\textstyle\int}}

\def\Ref#1{Ref.\cite{#1}}

\begin{document}
\allowdisplaybreaks[3]

\title{
Conservation laws and symmetries of \\
radial generalized nonlinear $p$-Laplacian\\
evolution equations
}

\author{
Elena Recio$^{1,2}$ \lowercase{\scshape{and}}
Stephen C. Anco$^1$
\\\lowercase{\scshape{
${}^1$Department of Mathematics and Statistics\\
Brock University\\
St. Catharines, ON L2S3A1, Canada}} \\\\
\lowercase{\scshape{
${}^2$Department of Mathematics\\
Faculty of Sciences, University of C\'adiz\\
Puerto Real, C\'adiz, Spain, 11510}}\\
}

\begin{abstract}
A class of generalized nonlinear $p$-Laplacian evolution equations is studied. 
These equations model radial diffusion-reaction processes in $n\geq 1$ dimensions, 
where the diffusivity depends on the gradient of the flow. 
For this class, 
all local conservation laws of low-order and all Lie symmetries are derived. 
The physical meaning of the conservation laws is discussed, 
and one of the conservation laws is used to show that the nonlinear equation 
can be mapped invertibly into a linear equation by a hodograph transformation
in certain cases. 
The symmetries are used to derive exact group-invariant solutions 
from solvable three-dimensional subgroups of the full symmetry group, 
which yields a direct reduction of the nonlinear equation to a quadrature. 
The physical and analytical properties of these exact solutions are explored, 
some of which describe moving interfaces and Green's functions. 
\end{abstract}

\maketitle

\begin{center}
email: 
elena.recio@uca.es, 
sanco@brocku.ca
\end{center}

\section{Introduction}

An interesting class of nonlinear diffusion-reaction equations is given by
the $p$-Laplacian evolution equations 
\begin{equation}
u_t = \kappa\nabla \cdot (|\nabla u|^{p-2}\nabla u) + ku^q, 
\quad
p>1,
\quad
q>0
\end{equation}
for $u(t,x)$ in $\Rnum\times \Rnum^n$.
Solutions of the initial-value problem exhibit several kinds of behaviour, 
such as finite-time blow up or finite-time extinction and moving interfaces, 
depending on the exponents $p$, $q$, and the sign of $k$ 
\cite{Dib,Gu,YuaLia,TiaMu,Zhao,LevParSer,Mes,LiuWan}. 
There are numerous physical applications of these equations, for example, 
filtration of non-Newtonian fluids, 
absorption in porous media,
combustion theory, 
turbulent gas flow,
glacial sliding, 
and continuum models of 
energy flows through discrete networks 
and vibrations of molecules 
\cite{Dib,SamGal,Zhao,WuZhao,ShoWal,PelRey}. 
Another prominent application occurs in image processing, 
where the $n=2$ equation with $k=0$ is used as an edge-enhancement model
\cite{PerMal}. 

This motivates studying a general nonlinear diffusion-reaction equation of the form 
\begin{equation}\label{nonlin_diffus_eqn}
u_t = \nabla \cdot (g(|\nabla u|) \nabla u) + f(u),
\quad
g'\neq 0,
\quad
f\neq 0
\end{equation}
which models diffusion processes 
in which the diffusivity $g$ depends on the gradient of $u$,
with a source/sink $f$ that depends on $u$.
Radial solutions $u=u(t,|x|)$ in $\Rnum\times\Rnum^n$ are of particular interest, 
due to the rotational symmetry of the equation \eqref{nonlin_diffus_eqn}.
These solutions satisfy the $n$-dimensional radial equation
\begin{equation}\label{nonlin_diffus_eqn_rad}
u_t = (g(|u_r|) +|u_r|g'(|u_r|))u_{rr} + (n-1)r^{-1}g(|u_r|)u_r + f(u),
\quad
g'\neq 0,
\quad
f\neq 0
\end{equation}
for $u(t,r)$ in $\Rnum\times\Rnum^+$ 
with the regularity conditions that 
the gradient
$\nabla u |_{x=0} = \hat{x} u_r |_{r=0}$ 
vanishes,
and the Hessian matrix
$\nabla^2 u |_{x=0} = 
((u_{rr} - r^{-1} u_r)\hat{x}\otimes\hat{x} + r^{-1} u_r I)|_{r=0}$
is proportional to the identity matrix $I$.
Note these regularity conditions can be formulated as 
\begin{equation}\label{regularity}
u_r\rightarrow 0 \text{ and } u_{rr} - r^{-1} u_r \rightarrow 0
\text{ for } r\rightarrow 0
\end{equation}
Also note that the diffusivity appearing in this radial equation \eqref{nonlin_diffus_eqn_rad}
is given by $h'(|u_r|)$ where $h(|u_r|) = |u_r|g(|u_r|)$. 

It will be natural to extend the radial equation \eqref{nonlin_diffus_eqn_rad}
by replacing $|u_r|$ with $u_r$,
and by allowing $n$ to have any value (not necessarily a non-negative integer). 
This yields a $1$-dimensional nonlinear diffusion-reaction equation 
\begin{equation}\label{nonlin_diffus_eqn_1D}
u_t = h'(u_r) u_{rr} + mr^{-1}h(u_r) + f(u),
\quad
h''\neq 0,
\quad
f\neq 0
\end{equation}
for $u(t,r)$ in $\Rnum\times\Rnum$, 
where
\begin{equation}\label{diffus_rel_1D}
h(u_r) = u_rg(u_r),
\quad
m=n-1
\end{equation}
(Note $r$ here is no longer a radial variable.)
When $h(u_r)$ is an odd function and $m$ is a non-negative integer, 
all solutions $u(t,r)$ of this equation \eqref{nonlin_diffus_eqn_1D}--\eqref{diffus_rel_1D} 
satisfying the regularity conditions \eqref{regularity}
on the half-line $0\leq r<\infty$ 
are solutions of the radial equation \eqref{nonlin_diffus_eqn_rad}
where 
\begin{equation}\label{g_rel}
g(|u_r|) = h(u_r)/u_r
\end{equation}
Hence, 
these solutions yield radial solutions $u(t,|x|)$ of the generalized $p$-Laplacian equation \eqref{nonlin_diffus_eqn} 
in $n=m+1$ dimensions. 
Radial solutions can also be obtained from solutions $u(t,r)$ 
whose radial derivative maintains a single sign,  
$u_r =|u_r| \geq 0$ or $u_r =-|u_r| \leq 0$ on the half-line $0\leq r<\infty$, 
since $g(|u_r|) = g(\pm u_r)= h(u_r)/u_r$ will be well-defined, 
without any restriction on $h(u_r)$. 
However, this sign condition will not hold in general for all solutions $u(t,r)$ of
the $1$-dimensional equation \eqref{nonlin_diffus_eqn_1D}--\eqref{diffus_rel_1D}. 

The purpose of the present paper is, firstly, 
to study the local conservation laws and the Lie symmetries admitted by 
the $1$-dimensional nonlinear diffusion-reaction equation \eqref{nonlin_diffus_eqn_1D} 
and its $n$-dimensional radial counterpart \eqref{nonlin_diffus_eqn_rad}, 
and secondly, 
to derive some exact group-invariant solutions by symmetry reduction. 

Local conservation laws are continuity equations that yield basic conserved quantities 
for all solutions \cite{Olv,BCA}. 
For nonlinear diffusion-reaction equations, 
they can be used to detect when the equation can be mapped invertibly into a linear equation \cite{AncBluWol}, 
and they allow checking the accuracy of numerical solution methods. 
More general conservation identities, such as total mass, 
provide physically important balance equations,
and they are used in obtaining estimates on $|u|$ or $|\nabla u|$
for classical solutions,
as well as in defining suitable norms for weak solutions.

Symmetries are transformations under which the whole solution space is mapped into itself \cite{Olv,BA,BCA}. 
For nonlinear diffusion-reaction equations, 
symmetries lead to exact group-invariant solutions
and play a role in defining invariant Sobolev norms.
Scaling symmetries are of special relevance,
as the critical nonlinearity power for blow-up behaviour or extinction behaviour of solutions
is typically singled out by scaling-invariance of a Sobolev norm.
Moreover, scaling transformation arguments give a means of
relating the behavior of solutions in different regimes, for instance,
solutions at short times with large initial data
can be scaled to long times with small initial data. 

There has been very little investigation of the symmetries and conservation laws 
for the class of equations \eqref{nonlin_diffus_eqn_1D} 
when the source term $f(u)$ is non-constant, 
as well as for the $n$-dimensional generalization \eqref{nonlin_diffus_eqn} of these equations. 
Recently, 
point symmetries have been classified in the special case of $n=1,2$ dimensions \cite{CheKinKov} 
and some group-invariant solutions have been obtained. 
However, no work at all appears to have been done on 
the radial nonlinear diffusion-reaction equation \eqref{nonlin_diffus_eqn_rad} for $n>1$
with a gradient-dependent diffusivity $h(|u_r|)$ 
when the source term $f(u)$ is non-constant. 
(Some other diffusion-reaction equations with a different form for the diffusivity
depending on the radial gradient $u_r$ 
and with a non-constant source term depending on $u$
have been studied in \Ref{JiQuYe}.)

A worthwhile remark is that a subclass of the reaction-diffusion equations \eqref{nonlin_diffus_eqn} 
given by $f(u)$ being linear in $u$ and $g(|\nabla u|)$ being a monomial in $|\nabla u|$
can be mapped by a point transformation 
into nonlinear filtration equations 
$u_t = \nabla \cdot (|\nabla u|^{p-2}) \nabla u)$. 
There is a sizeable literature on symmetries and conservation laws of these equations, 
especially in the case $n=1$ \cite{IbrAkhGaz,Zha}
where they can be transformed via $u_x=v$ 
into standard diffusion equations $v_t = g(v)_{xx}$ with a non-gradient diffusivity
which have been investigated extensively 
(see, for example, \Ref{Dor,DorVir,Qu,CheSerRas} and references therein). 

The present paper is organized as follows. 

In section \ref{conslaws_sec},
we derive all local conservation laws of low-order 
for the $1$-dimensional nonlinear diffusion-reaction equation \eqref{nonlin_diffus_eqn_1D}--\eqref{diffus_rel_1D} 
with a general gradient diffusivity, 
by using the general method of multipliers 
\cite{AncBlu97,AncBlu02a,AncBlu02b,Anc-review}. 
We also discuss the physical meaning of these conservation laws. 
This classification of low-order conservation laws is new. 
From one of the conservation laws, 
we show that the nonlinear equation \eqref{nonlin_diffus_eqn_1D}--\eqref{diffus_rel_1D} 
can be mapped invertibly into a linear equation by a hodograph transformation 
in certain cases when (and only when) the radial diffusivity has the form 
$h'(u_r) =\kappa/(\alpha+u_r{})^2$.

In section \ref{symms_sec},
we derive all of the Lie symmetries for the $1$-dimensional nonlinear diffusion-reaction equation \eqref{nonlin_diffus_eqn_1D}--\eqref{diffus_rel_1D} 
with a general gradient diffusivity. 
We also work out the corresponding symmetry transformations,
and we present a complete classification of the admitted maximal symmetry groups. 
The results obtained for all cases $m\neq0$ are new. 

Next, in section \ref{lift_sec}, 
we present the local conservation laws and Lie symmetries of 
the radial nonlinear diffusion-reaction equation \eqref{nonlin_diffus_eqn_rad}
with a general gradient diffusivity. 
We show that the same hodograph transformation 
which maps the $1$-dimensional equation into a linear equation
also works to map the radial equation into a linear equation. 
All of these results are new. 

In section \ref{solns_sec}, 
we use the Lie symmetries to derive exact group-invariant solutions of the radial nonlinear diffusion-reaction equation \eqref{nonlin_diffus_eqn_rad}
for special forms of the gradient diffusivity. 
The solutions are obtained in an explicit form by using 
solvable three-dimensional subgroups of the symmetry group,
which yields a direct reduction of the equation \eqref{nonlin_diffus_eqn_rad} 
to a quadrature. 
We examine some of the physical and analytical properties of these exact solutions. 
In particular, 
one family of solutions describes moving interfaces (governed by a Stephan condition),
and another family of solutions describes Green's functions (involving a Dirac delta source term). 
For comparison, 
some interesting similar solutions that are not group-invariant can be found in 
Ref.\cite{Gal}.  

In section \ref{remarks_sec},
we make some final concluding remarks.

\section{Low-order conservation laws}
\label{conslaws_sec}

A \emph{local conservation law} of the $1$-dimensional nonlinear diffusion-reaction equation \eqref{nonlin_diffus_eqn_1D}--\eqref{diffus_rel_1D} 
is a continuity equation 
\begin{equation}\label{conslaw}
(D_t T + D_r X )|_\Esp=0
\end{equation}
holding on the whole solution space $\Esp$ of the equation \eqref{nonlin_diffus_eqn_1D}--\eqref{diffus_rel_1D},
where the conserved density $T$ and the radial flux $X$
are functions of $t$, $r$, $u$, and radial derivatives of $u$, 
with time derivatives of $u$ being eliminated from $T$ and $X$ 
through equation \eqref{nonlin_diffus_eqn_1D}. 
Here $D_t$ and $D_r$ denote total derivatives with respect to $t$ and $r$. 

On any domain $\Omega\in\Rnum$ with endpoints $\partial\Omega=\{a,b\}$, 
the radial integral of the conserved density
\begin{equation}\label{conserved_quant}
\C [u] = \int_{a}^{b} T \,dr 
\end{equation}
satisfies the integral continuity equation 
\begin{equation}\label{integral_conslaw}
\frac{d}{dt} \C[u] = - X\Big|_{a}^{b}
\end{equation}
which relates the rate of change of the integral quantity $\C[u]$ on the domain
to the net radial flux through the endpoints. 
This integral quantity $\C[u]$ will be conserved for solutions $u(t,r)$ such that 
the net radial flux is zero. 
Conversely,
any such conserved quantity $\C[u]$
arises from a local conservation law \eqref{conslaw}.

Two local conservation laws are \emph{locally equivalent} 
if they yield the same conserved quantity
$\C [u]$ up to boundary terms. 
This happens iff their conserved densities $T$ 
differ by a total radial derivative $D_r \Theta$ of some function $\Theta$ 
of $t$, $r$, $u$, and radial derivatives of $u$, 
and correspondingly, their radial fluxes $X$ 
differ by a total time derivative $-D_t \Theta$ evaluated on $\Esp$. 
A conservation law is thereby called
\emph{locally trivial}
if
\begin{equation}
T = D_r \Theta, 
\quad
X = - (D_t \Theta)|_\Esp
\end{equation}
Thus, locally equivalent conservation laws differ by locally trivial conservation laws.
The set of all admitted local conservation laws forms a vector space,
in which the set of locally trivial conservation laws is a subspace. 

Each local conservation law \eqref{conslaw}
has an equivalent \emph{characteristic form}
that holds as an identity off of the solution space $\Esp$ of 
the radial diffusion-reaction equation \eqref{nonlin_diffus_eqn_1D}--\eqref{diffus_rel_1D}. 
This identity provides the basis for setting up a general method using multipliers 
to find all local conservation laws. 
In particular, 
the multiplier method can be formulated as a kind of adjoint of the standard method
for finding symmetries 
\cite{AncBlu97,Anc-review}. 

The characteristic form of a local conservation law is obtained from the relation between 
the total time derivative on $\Esp$ and off $\Esp$:
\begin{equation}
\begin{split}
D_t =
& D_t|_{\Esp}
+ \left(u_t - h'(u_r) u_{rr} - mr^{-1}h(u_r) - f(u)\right) \partial_{u}
\\& \qquad 
+ D_r\left(u_t - h'(u_r) u_{rr} - mr^{-1}h(u_r) - f(u)\right)\partial_{u_r} 
+ \cdots
\end{split}
\end{equation}
Substitution of this relation into a conservation law \eqref{conslaw} yields 
\begin{equation}
\begin{aligned}
D_t T + D_r X = & 
\left(u_t - h'(u_r) u_{rr} - mr^{-1}h(u_r) - f(u)\right) T_{u}
\\&\qquad
+ D_r\left(u_t - h'(u_r) u_{rr} - mr^{-1}h(u_r) - f(u)\right) T_{u_r} 
+\cdots
\end{aligned}
\end{equation}
which holds identically. 
Integration by parts on the terms on the right hand side then gives 
\begin{equation}\label{char_eqn}
D_t T + D_r(X - \Psi) =
\left(u_t - h'(u_r) u_{rr} - mr^{-1}h(u_r) - f(u)\right) Q
\end{equation}
which defines the characteristic equation, 
where
\begin{equation}
\begin{split}
\Psi = & 
\left(u_t - h'(u_r) u_{rr} - mr^{-1}h(u_r) - f(u)\right) \hat E_{u_r}(T)
\\& \qquad 
+ D_r\left(u_t - h'(u_r) u_{rr} - mr^{-1}h(u_r) - f(u)\right) \hat E_{u_{rr}}(T)
+\cdots
\end{split}
\end{equation}
vanishes on $\Esp$, 
and where
\begin{equation}\label{mult}
Q = \hat E_{u}(T)
\end{equation}
is a function of $t$, $r$ $u$, and radial derivatives of $u$. 
Here
$\hat E_w = \partial_w - D_r \partial_{w_r} + D_r^2 \partial_{w_{rr}} - \cdots$
denotes the spatial Euler operator \cite{BCA,Olv}
with respect to a variable $w$.
This operator has the important property that 
a function $F(t,r,w,w_r,w_{rr},\ldots)$ is annihilated by $\hat E_w$
iff the function is a total radial derivative,
$F = D_r G(t,r,w,w_r,w_{rr})$.

When evaluated on the solution space $\Esp$, 
the characteristic equation \eqref{char_eqn} 
reduces to the local conservation law \eqref{conslaw},
since the flux term $\Psi|_\Esp=0$ is trivial. 
The function $Q$ is called the \emph{multiplier} (or \emph{characteristic}) 
of the conservation law. 

Most importantly, 
the relation \eqref{mult} between $Q$ and $T$ shows that 
all locally equivalent conservation laws have the same multiplier. 
In particular,
if a conserved density is locally trivial, $T = D_r \Theta$, 
then its multiplier vanishes, 
since $E_u(D_r \Theta) = 0$.
Conversely,
if a multiplier is trivial, $Q = E_{u}(T)=0$, 
this implies that $T = D_r \Theta$ is a locally trivial conserved density, 
whereby the characteristic equation becomes 
$D_t D_r\Theta + D_r(X-\Psi) = 0$ 
giving $X = \Psi-D_t \Theta$ 
(after an arbitrary function of $t$ is absorbed into $\Theta$), 
which is a locally trivial radial flux. 

Thus, through the characteristic equation for conservation laws, 
there is a one-to-one correspondence between multipliers and conserved densities
(up to local equivalence). 
This result holds more generally for any evolution PDE
\cite{Olv,Anc-review}

In general, 
a function $Q$ of $t$, $r$ $u$, and radial derivatives of $u$ 
is a multiplier for a local conservation law iff 
the product of $Q$ and the radial diffusion-reaction equation \eqref{nonlin_diffus_eqn_1D} 
is a total divergence with respect to $t$ and $r$. 
All multipliers can be determined by the property that 
a function $h(t,r,u,u_r,u_t,u_{rr},u_{tr},u_{tt},\dots)$ is such a divergence 
iff $E_u(h)=0$ holds identically, 
where 
\begin{equation}\label{Eop}
E_w = \partial_w - D_t\partial_{w_t} - D_r \partial_{w_r} 
+ D_t^2 \partial_{w_{tt}} + D_tD_r \partial_{w_{tr}} + D_r^2 \partial_{w_{rr}} 
- \cdots
\end{equation}
denotes the Euler operator with respect to a variable $w$
\cite{Olv,BCA,Anc-review}. 
(Note that the restriction of $E_w$ acting on functions without time derivatives is $\hat E_w$.)
This yields the determining equation 
$E_u\big((u_t - h'(u_r) u_{rr} - mr^{-1}h(u_r) - f(u))Q\big) =0$. 
Since $Q$ does not involve $u_t$ and radial derivatives of $u_t$, 
the determining equation splits with respect to these variables
into an overdetermined linear system of equations
\begin{gather}
-(D_t Q)|_\Esp -D_r(h' D_r Q) +D_r(mr^{-1}h'Q) -f' Q =0
\label{adj_symm_det_eqn}\\
Q_u -\hat E_u(Q) =0,
\quad
Q_{u_r} +\hat E^{(1)}_{u_r}(Q) =0,
\quad
Q_{u_{rr}} - \hat E^{(2)}_{u_{rr}}(Q) =0,
\quad
\text{etc.}
\label{helmholtz_eqns}
\end{gather}
which is the standard determining system for multipliers. 
Here 
$\hat E^{(1)}_w = \partial_w - \binom{2}{1}D_r \partial_{w_r} + \binom{3}{1}D_r^2 \partial_{w_{rr}} - \cdots$
and 
$\hat E^{(2)}_w = \partial_w - \binom{3}{2}D_r \partial_{w_r} + \binom{4}{2}D_r^2 \partial_{w_{rr}} - \cdots$
denote higher spatial Euler operators
\cite{Olv,Anc-review}. 
It is useful to note that equation \eqref{adj_symm_det_eqn} turns out to be the adjoint of the determining equation for symmetries,
and that equation \eqref{helmholtz_eqns} represents Helmholtz conditions that are necessary and sufficient for $Q$ to have the variational form \eqref{mult}. 

The solutions $Q(t,r,u,u_r,u_{rr},\ldots)$ of this determining system \eqref{adj_symm_det_eqn}--\eqref{helmholtz_eqns}
yield all conservation law multipliers for the nonlinear diffusion-reaction equation \eqref{nonlin_diffus_eqn_1D}--\eqref{diffus_rel_1D}. 
For nonlinear diffusion-reaction equations in general, 
the local conservation laws of physical interest typically arise from low-order multipliers
\cite{Anc16a,Anc-review}. 
The general form for \emph{low-order multipliers} $Q$ 
in terms of $u$ and derivatives of $u$ 
is given by those variables that can be differentiated to obtain 
a leading derivative of the given nonlinear diffusion-reaction equation. 
The leading derivatives of equation \eqref{nonlin_diffus_eqn_1D} 
consist of $u_t$ and $u_{rr}$. 
Clearly, $u_t$ can be obtained by differentiation of $u$, 
while $u_{rr}$ can be obtained by differentiation of $u$ and $u_r$. 
This determines 
\begin{equation}\label{low_order}
Q(t,r,u,u_r)
\end{equation}
as the general form for a low-order multiplier for the nonlinear diffusion-reaction equation \eqref{nonlin_diffus_eqn_1D}--\eqref{diffus_rel_1D}. 

For low-order multipliers \eqref{low_order}, 
the determining system \eqref{adj_symm_det_eqn}--\eqref{helmholtz_eqns} reduces to 
the system 
\begin{subequations}\label{mult_det_sys}
\begin{align}
& Q_{u_r}=0\\
& (Q_r - mr^{-1}Q) h''+Q_u (u_r h''+2h')= 0 \\
& Q_t + (f Q)_u +m( r^{-1}Q_u (h -u_r h') -(r^{-1}Q)_r h' )
+( Q_{rr} +2Q_{ru}u_r +Q_{uu}u_r{}^2 ) h' =0
\end{align}
\end{subequations}
for $Q(t,r,u,u_r)$, $h(u_r)$, $f(u)$, $m$. 
Note, from equation \eqref{nonlin_diffus_eqn_1D}, 
we have the conditions $h''\neq 0$ and $f\neq 0$, 
which respectively imply the diffusivity is non-constant and the source/sink is non-zero. 

This determining system \eqref{mult_det_sys} can be split with respect to $u_r$,
by using $h'\neq 0$ and $h''\neq 0$ as pivots and then differentiating by $u_r$. 
To derive the split system and obtain its complete set of solutions, 
we have used the steps outlined in the appendix. 

The final tree of solution cases is listed in table~\ref{table_multr}. 
The first two solution cases each describe 
one-dimensional linear sets of multipliers $\{Q(t,r,u)\}$.
These two cases are mutually exclusive. 
The last two solution cases describe 
infinite-dimensional linear sets of multipliers $\{Q(t,r,u)\}$
which depend on an arbitrary solution $\psi(t,z)$ of the linear PDE
\begin{equation}\label{linear_eqn}
\psi_t +\kappa \psi_{zz}+(f(z)\psi)_z = 0
\end{equation}
These two cases also are mutually exclusive. 
Note that there is no solution case in which $f(u)$, $h(u_r)$, and $m$ all are arbitrary. 

\begin{table}[htb]
\centering
\caption{Case tree of low-order multipliers}
\label{table_multr}
\begin{tabular}{c|ccc}
\hline
$Q(t,r,u,u_r)$ & $h(u_r)$ & $f(u)$ & $m$ 
\\
\hline
\hline
$r^m \exp(-kt)$
&
&
$a+ku$
&
\\
\hline
$r^m \exp(-p((p\kappa-a)t +\alpha r+u))$
&
$-\kappa/(\alpha+u_r)$
&
$a+k\exp(pu)$
&
\\
&
&
$k,p \neq0$
&
\\
\hline
\hline
$r^m \psi(t,u)$
&
$-\kappa/u_r$
&
&
\\
\hline
$r^m \psi(t,u+\alpha r)$
&
$-\kappa/(\alpha+u_r)$ 
&
$a$
&
\\
$\alpha\neq 0$
&
&
&
\\
\hline
\end{tabular}
\end{table}

Each low-order multiplier $Q$ determines 
a corresponding conserved density $T$ and radial flux $X$ 
through the characteristic equation \eqref{char_eqn}. 
Note that, since $Q_{u_r}=0$, the multiplier relation \eqref{mult} directly implies 
$T$ is a function only of $t,r,u$,
while the characteristic equation \eqref{char_eqn} then implies  
$X$ is a function only of $t,r,u,u_r$. 
Hence, all low-order local conservation laws have the general form 
\begin{equation}\label{low_order_TX}
T(t,r,u),
\quad
X(t,r,u,u_r)
\end{equation}
There are two general methods to obtain $T$ and $X$ explicitly in terms of $Q$
\cite{BCA,Anc-review}. 
One method consists of first splitting the characteristic equation \eqref{char_eqn} 
with respect to $u_t, u_{tt}, u_{tr}, u_{rr}$, 
and then integrating the resulting linear system
\begin{subequations}
\begin{align}
& T_u -Q =0\\
& X_{u_r} +h' T_u =0\\
& T_t + X_r +  u_r X_u +(mr^{-1}h+f)T_u =0
\end{align}
\end{subequations}
A second method uses a homotopy integral formula, 
or an equivalent algebraic scaling formula \cite{Anc03,Anc-review}, 
which can be straightforwardly derived from this linear system. 

By applying either of these two methods, 
we have the following classification result, 
which is arranged by generality of the diffusivity $h'(u_r)$.

\begin{thm}\label{cons_laws_thm}
(i) The low-order local conservation laws admitted by 
the $1$-dimensional nonlinear diffusion-reaction equation \eqref{nonlin_diffus_eqn_1D}--\eqref{diffus_rel_1D} 
for arbitrary diffusivity $h'(u_r)$ 
are given by 
\begin{subequations}\label{cons_law1}
\begin{align}
& f(u) = a+ku 
\\
&\begin{aligned}
& T_1=r^m \exp(-kt) u
\\
& X_1 = \begin{cases}
- r^m \exp(-kt) (h(u_r) + a r/(m+1)), & m\neq -1\\
- \exp(-kt) (r^{-1}h(u_r) + a \ln|r| ), & m=-1
\end{cases}
\end{aligned}
\end{align}
\end{subequations}
modulo a trivial conserved density and radial flux.
\newline
(ii) The $1$-dimensional nonlinear diffusion-reaction equation \eqref{nonlin_diffus_eqn_1D}--\eqref{diffus_rel_1D} 
admits additional low-order local conservation laws only for diffusivities $h'(u_r)$ 
given by 
\begin{equation}\label{extra_cons_law}
h(u_r)= -\kappa/(\alpha+u_r)
\end{equation}
These conservation laws consist of 
\begin{subequations}\label{cons_law2}
\begin{flalign}
& \alpha=0,
\quad
f(u) \text{ arbitrary }&
\label{case2}\\
& \begin{aligned}
& T_{2,\infty}= r^m \phi(t,u) \\
& X_{2,\infty}= \kappa r^m \frac{\phi_z(t,u)}{u_r} \\
\end{aligned}&
\end{flalign}
\end{subequations}
\begin{subequations}\label{cons_law3}
\begin{flalign}
&\alpha\neq0,
\quad
f(u) = a+k\exp(pu),
\quad
p,k\neq 0&
\\
& \begin{aligned}
& T_{3}= r^m \exp(-p((p\kappa-a) t +\alpha r +u)) \\
& X_{3}= -\exp(p(a-p\kappa) t)\left( \frac{p\kappa  r^m \exp(-p(\alpha r+u))}{\alpha+u_r}
+k\alpha^{-m-1} p^{-m}\Gamma(m+1,p\alpha r) \right)
\end{aligned}&
\end{flalign}
\end{subequations}
\begin{subequations}\label{cons_law4}
\begin{flalign}
& \alpha\neq 0,
\quad
f(u) = a&
\label{case4}\\
& \begin{aligned}
& T_{4,\infty}= r^m \phi(t,\alpha r+u) \\
& X_{4,\infty}= \kappa r^m \frac{\phi_z(t,\alpha r+u)}{\alpha+u_r} \\
\end{aligned}&
\end{flalign}
\end{subequations}
modulo a trivial conserved density and radial flux,
where $\phi(t,z)$ satisfies the linear PDE 
\begin{equation}\label{alt_linear_eqn}
\phi_t +\kappa\phi_{zz}+f(z)\phi_{z} = 0
\end{equation}
and where $\Gamma(q,z)$ denotes the incomplete gamma function. 
\end{thm}

Note this classification shows that there are no special dimensions $n=m+1$ 
in which extra low-order conservation laws are admitted. 
Also, note that the special diffusivity cases have been arranged to be mutually exclusive.

We will now discuss the physical meaning and the main analytical properties 
of the low-order conservation laws \eqref{cons_law1}--\eqref{cons_law4}
expressed as continuity equations \eqref{integral_conslaw}
for integral quantities \eqref{conserved_quant}. 

We first consider the two conservation laws \eqref{cons_law1} and \eqref{cons_law3}, 
which do not involve the function $\phi$. 
Let $\Omega\in\Rnum$ be either the full-line $(-\infty,+\infty)$ or the half-line $[0,+\infty)$,
with $\partial\Omega$ being the endpoints $\{-\infty,\infty\}$ or $\{0,\infty\}$. 

For conservation law \eqref{cons_law1}, 
the integral quantity is given by 
\begin{equation}\label{cons_quantity1}
\C_1[u] = e^{-kt} \int_{\Omega} u(t,r)r^m \, dr
\end{equation}
which is finite for solutions $u(t,r)$ such that 
$r^{m+1}u\rightarrow 0$ as $r\rightarrow \partial\Omega$. 
This integral quantity \eqref{cons_quantity1} will be conserved iff 
$a=f(0)=0$ and $r h(u_r)\rightarrow 0$ as $r\rightarrow \partial\Omega$. 
Under these conditions, 
the mass integral
\begin{equation}\label{mass}
M(t) = \int_{\Omega} u(t,r) r^m\, dr = e^{kt} M_0
\end{equation}
is an exponential function of $t$, 
where 
\begin{equation}
M_0 = M(0)=\C_1[u] =\const
\end{equation}
is the conserved quantity \eqref{cons_quantity1}. 
Consequently, 
as $t \rightarrow \infty$,
$M(t) \rightarrow \infty$
when $k > 0$
and
$M(t) \rightarrow 0$
when $k < 0$,
corresponding to when the term $f(u)=ku$ 
in the $1$-dimensional diffusion-reaction equation \eqref{nonlin_diffus_eqn_1D}--\eqref{diffus_rel_1D} 
acts as either a source or a sink.

For conservation law \eqref{cons_law3}, 
the integral quantity is given by 
\begin{equation}\label{cons_quantity3}
\C_3[u] = e^{-p^2\kappa t} \int_{\Omega} e^{-p(u(t,r)+\alpha r )} r^m \, dr
\end{equation}
This integral is finite for solutions $u(t,r)$ such that 
$r^{m+1}e^{-p(\alpha r+u )}\rightarrow 0$ as $r\rightarrow \partial\Omega$. 
The resulting asymptotic behaviour of $u$ can be shown to imply 
$h(u_r)= -\kappa/(\alpha +u_r) \sim O(r)$,
which is sufficient to show that the term 
$p\kappa  r^m e^{-p(\alpha r+u)}/(\alpha+u_r)$ in the radial flux 
vanishes as $r\rightarrow \partial\Omega$. 
However, 
the other term $a\alpha^{-m-1} p^{-m}\Gamma(m+1,p\alpha r)$ in the radial flux 
vanishes only for  $r\rightarrow +\infty$, 
since $\Gamma(m+1,p\alpha r)\rightarrow \Gamma(m+1)$ for $r\rightarrow 0$,
and $\Gamma(m+1,p\alpha r)$ diverges for $r\rightarrow -\infty$. 
As a result, 
$\C_3[u]$ is conserved only if $a=f(0)=0$. 
In this case, 
the integral 
\begin{equation}
S(t) = \int_{0}^{\infty} e^{-p(u(t,r)+\alpha r )} r^m \, dr
= e^{p^2\kappa t} S_0
\end{equation}
is an exponential function of $t$, 
where 
\begin{equation}
S_0 = S(0)=\C_3[u] =\const
\end{equation}
is the conserved quantity \eqref{cons_quantity3}. 
Hence, 
$S(t) \rightarrow \infty$
as $t \rightarrow \infty$
if $\kappa>0$, 
which holds when the radial diffusivity $h'(u_r)= \kappa/(\alpha +u_r)^2$ is non-negative. 

The remaining two conservation laws \eqref{cons_law2} and \eqref{cons_law4},  
both of which involve the function $\phi$, 
are related to a hodograph transformation, 
as we will discuss next.

\subsection{Hodograph transformation} 

The conservation laws \eqref{cons_law2} and \eqref{cons_law4}
are singled out because the function $\phi$ satisfies a linear PDE \eqref{alt_linear_eqn}, 
whereby each conservation law represents an infinite family of continuity equations. 
According to a general mapping theorem \cite{AncBluWol},
this indicates that the $1$-dimensional nonlinear diffusion-reaction equation \eqref{nonlin_diffus_eqn_1D}--\eqref{diffus_rel_1D} 
can be mapped invertibly into a linear diffusion equation 
by a point transformation
in the cases \eqref{case2} and \eqref{case4} 
for which the two conservation laws respectively hold. 

We will discuss the case \eqref{case2} first. 
The nonlinear diffusion-reaction equation \eqref{nonlin_diffus_eqn_1D}--\eqref{diffus_rel_1D} 
in this case is given by 
\begin{equation}\label{nonlin_diffus_eqn_1D_hodogr2}
u_t = \kappa \frac{u_{rr}-mr^{-1}u_r}{u_r{}^2}  + f(u)
\end{equation}
There are four steps in deriving the point transformation 
$(t,r,u)\rightarrow (\tilde{t},\tilde{r},\tilde{u})$ 
that linearizes this equation, 
following the method shown in \cite{AncBluWol}. 
For the first step, 
the new independent variables $(\tilde{t},\tilde{r})$
are given by the arguments of the function $\phi(t,u)$
appearing in the conservation law \eqref{cons_law2}. 
This yields 
\begin{equation}
\tilde{t} = t, 
\quad
\tilde{r} = u
\end{equation}
Note this part of the transformation, 
$(t,r)\rightarrow (\tilde{t},\tilde{r})$, 
has the Jacobian $J = \det \left(D(\tilde{t},\tilde{r})/D(t,r)\right) =u_r$.
For the second step, 
the characteristic equation \eqref{char_eqn} 
for the conservation law \eqref{cons_law2} is expressed 
as an identity in terms of the function $\phi(t,u)$, 
\begin{equation}\label{char_eqn2}
\begin{aligned}
& D_t( r^m \phi ) + D_r( \kappa r^m \phi_u/u_r ) 
- \big(u_t + \kappa(u_{rr} - m r^{-1} u_r)/u_r{}^2  - f(u)\big)r^m \phi_u
\\&
= r^m \big( \phi_t - \kappa \phi_{uu} + f(u) \phi_u \big)
\\& 
= D_r \left( \smallint r^m\,dr \big( \phi_t - \kappa \phi_{uu} + f(u) \phi_u \big) \right)
- \smallint r^m\,dr u_r \big( \psi_t - \kappa \psi_{uu} + (f(u) \psi)_u \big)
\end{aligned}
\end{equation}
where $\psi(t,u)=\phi_u(t,u)$ is the function appearing in the multiplier 
for this conservation law in table~\ref{table_multr}. 
When $\phi$ satisfies the linear PDE \eqref{alt_linear_eqn}, 
the function $\psi$ satisfies the linear PDE  \eqref{linear_eqn}
which appears on the righthand side of the identity \eqref{char_eqn2}. 
For the next step, 
the new dependent variable $\tilde{u}$ is obtained by 
equating the corresponding multiplier $-u_r\int r^m\,dr$ 
to the expression $-\tilde{u} J$, 
which yields
\begin{equation}\label{new_u}
\tilde{u} =\int r^m\,dr 
=\begin{cases}
\tfrac{1}{m+1} r^{m+1},  & m\neq -1\\
\ln|r|, & m=-1
\end{cases}
\end{equation}
For the final step, 
the adjoint of the linear PDE \eqref{linear_eqn} for $\psi(t,u)$ 
gives the linear diffusion equation satisfied by $\tilde{u}(\tilde{t},\tilde{r})$, 
\begin{equation}\label{linear_diffus_eqn}
\tilde{u}_{\tilde{t}} = \kappa \tilde{u}_{\tilde{r}\tilde{r}} + f(\tilde{r})\tilde{u}_{\tilde{r}} 
\end{equation}

As a check,
the point transformation
\begin{equation}
(t,r,u) \rightarrow (\tilde{t},\tilde{r},\tilde{u})
= (t,u,\tfrac{1}{n} r^n)
\end{equation}
has the prolongation
\begin{equation}
\tilde{u}_{\tilde{t}} = - r^m u_t/u_r, 
\quad
\tilde{u}_{\tilde{r}} = r^m/u_r, 
\quad
\tilde{u}_{\tilde{r}\tilde{r}} = - r^m (u_{rr} -m r^{-1} u_r)/u_r{}^3
\end{equation}
giving
\begin{equation}
u_t  = - \tilde{u}_{\tilde{t}}/\tilde{u}_{\tilde{r}}, 
\quad
u_r = r^m/ \tilde{u}_{\tilde{r}}, 
\quad
u_{rr}  = -r^m(r^m \tilde{u}_{\tilde{r}\tilde{r}} -m r^{-1} \tilde{u}_{\tilde{r}})/\tilde{u}_{\tilde{r}}{}^3 
\end{equation}
under which the nonlinear diffusion-reaction equation \eqref{nonlin_diffus_eqn_1D_hodogr2}
is seen to be equivalent to the linear diffusion equation \eqref{linear_diffus_eqn}. 

We will next discuss the case \eqref{case4}. 
The nonlinear diffusion-reaction equation \eqref{nonlin_diffus_eqn_1D}--\eqref{diffus_rel_1D} 
in this case is given by 
\begin{equation}\label{nonlin_diffus_eqn_1D_hodogr4}
u_t = \kappa \frac{u_{rr}-mr^{-1}(\alpha +u_r)}{(\alpha+u_r)^2}  + a
\end{equation}
We apply the previous four steps again. 
First, 
from the conservation law \eqref{cons_law4}, 
the new independent variables $(\tilde{t},\tilde{r})$ in the point transformation 
are given by the arguments of the function $\phi(t,\alpha r+u)$,  
\begin{equation}
\tilde{t} = t, 
\quad
\tilde{r} = \alpha r + u
\end{equation}
where $J = \det \left(D(\tilde{t},\tilde{r})/D(t,r)\right) =\alpha +u_r$
is the Jacobian. 
Second, 
the characteristic equation for the conservation law \eqref{cons_law4} 
is expressed as an identity 
\begin{equation}\label{char_eqn4}
\begin{aligned}
& D_t( r^m \phi ) + D_r( \kappa r^m \phi_z/(\alpha+u_r) ) 
- \big(u_t + \kappa(u_{rr} - m r^{-1}(\alpha +u_r))/(\alpha+u_r)^2  - a\big)r^m \phi_z
\\&
= D_r \left(\smallint r^m\,dr  \big( \phi_t - \kappa \phi_{zz} + a\phi_z \big) \right)
- \smallint r^m\, dr (\alpha +u_r) \big( \psi_t - \kappa \psi_{zz} + a\psi_z \big)
\end{aligned}
\end{equation}
where $\psi(t,z)=\phi_z(t,z)$ is the function appearing in the multiplier 
for this conservation law in table~\ref{table_multr},
with $z=\alpha r +u$. 
When $\phi$ satisfies the linear PDE \eqref{alt_linear_eqn} with $f(u)=a$, 
the function $\psi$ satisfies the linear PDE  \eqref{linear_eqn}
which appears on the righthand side of the identity \eqref{char_eqn4}. 
Third, 
from the multiplier relation $-(\alpha+u_r) \smallint r^m\, dr=-\tilde{u} J$,
the new dependent variable $\tilde{u}$ is given by the expression \eqref{new_u}. 
Finally, the adjoint of the linear PDE \eqref{linear_eqn} for $\psi$ 
gives the linear diffusion equation \eqref{linear_diffus_eqn}
satisfied by $\tilde{u}(\tilde{t},\tilde{r})$, 
with $f(u)=a$. 

The classification of low-order conservation laws in Theorem~\ref{cons_laws_thm}
combined with the general mapping theorem in \cite{AncBluWol} 
now establishes the following result. 

\begin{prop}\label{hodograph}
The $1$-dimensional nonlinear diffusion-reaction equation \eqref{nonlin_diffus_eqn_1D}--\eqref{diffus_rel_1D} 
can be mapped into a linear diffusion equation by a point transformation iff
the diffusivity has the form $h'(u_r)= \kappa/(\alpha +u_r)^2$
and the source/sink has the form $f(u)=a$ when $\alpha\neq0$ 
and $f(u)$ arbitrary when $\alpha=0$. 
The linear diffusion equation is given by 
$\tilde{u}_{\tilde{t}} = \kappa \tilde{u}_{\tilde{r}\tilde{r}} + f(\tilde{r})\tilde{u}_{\tilde{r}}$
where $(t,r,u) \rightarrow (\tilde{t},\tilde{r},\tilde{u})= (t,\alpha r +u,\smallint r^m\, dr)$
is the point transformation. 
\end{prop}

The special case $m=0$ of this hodograph transformation, 
which applies to a reaction-diffusion equation with a gradient diffusivity in one spatial dimension, 
was first derived in \Ref{CheKinKov}. 
It is interesting that the same transformation extends to higher spatial dimensions.

\section{Lie Symmetries}
\label{symms_sec}

An \emph{infinitesimal point symmetry} of 
$1$-dimensional nonlinear diffusion-reaction equation \eqref{nonlin_diffus_eqn_1D}--\eqref{diffus_rel_1D} 
is a generator of the form
\begin{equation}\label{point_symm}
\X=
\tau(t,r,u)\partial_t + \xi(t,r,u)\partial_r +\eta(t,r,u)\partial_u
\end{equation}
whose prolongation leaves invariant
the solution space $\Esp$ of equation \eqref{nonlin_diffus_eqn_1D}--\eqref{diffus_rel_1D},
\begin{equation}
\pr\X \left(u_t -h'(u_r)u_{rr} -m r^{-1}h(u_r) -f(u)\right)|_{\Esp} = 0
\end{equation}
Exponentiation of the point symmetry generator \eqref{point_symm}
produces a one-parameter symmetry transformation group
\begin{equation}\label{trans_group}
(t,r,u) \rightarrow
(\tilde{t},\tilde{r},\tilde{u}) 
= \exp (\epsilon \X) (t,r,u)
\end{equation}
where $\epsilon$ 
is the group parameter such that
$(\tilde{t},\tilde{r},\tilde{u})|_{\epsilon=0}=(t,r,u)$
is the identity transformation.
The explicit form of this symmetry group
\eqref{trans_group}
can be obtained by solving the ODE system
\begin{equation}\label{ODEs_symm_group}
\frac{\partial \tilde{t}}{\partial \epsilon} = \tau (\tilde{t},\tilde{r},\tilde{u}),
\quad
\frac{\partial \tilde{r}}{\partial \epsilon} = \xi (\tilde{t},\tilde{r},\tilde{u}),
\quad
\frac{\partial \tilde{u}}{\partial \epsilon} = \eta (\tilde{t},\tilde{r},\tilde{u})
\end{equation}
with the initial conditions
\begin{equation}\label{ics_symm_group}
\tilde{t} |_{\epsilon=0} = t, 
\quad
\tilde{r} |_{\epsilon=0} = r, 
\quad
\tilde{u} |_{\epsilon=0} = u 
\end{equation}

The infinitesimal action of a point symmetry 
\eqref{trans_group} on solutions $u(t,r)$ of
$1$-dimensional nonlinear diffusion-reaction equation \eqref{nonlin_diffus_eqn_1D}--\eqref{diffus_rel_1D} 
is given by
\begin{equation}\label{inf_action}
\begin{aligned}
u(t,r) \rightarrow 
\tilde{u}(t,r) =  u(t,r) 
& + \epsilon \left(\eta(t, r,u(t,r)) -\tau(t,r,u(t,r)) u_t(t,r) \right. 
\\ 
& \qquad \left. - \xi(t,r,u(t,r)) u_r(t,r) \right) + O(\epsilon^2)
\end{aligned}
\end{equation}
which corresponds to a generator
\begin{equation}\label{Xchar_form}
\hat{\X}=P \partial_u, 
\quad 
P=\eta-\tau u_t-\xi u_r
\end{equation}
This is called the \emph{characteristic form}
of the infinitesimal point symmetry.
The invariance condition then takes the form
\begin{equation}\label{sym_det_eqn}
\begin{aligned}
0 =  &  \pr \hat{\X} \left(u_t -h'(u_r)u_{rr} - m r^{-1}h(u_r) -f(u)\right)|_{\Esp} 
\\
= & \left. \left( D_t P 
-h'(u_r) D_r^2 P 
-(m r^{-1} h'(u_r) +h''(u_r) u_{rr})D_r P -f'(u) P  \right) \right|_{\Esp}
\end{aligned}
\end{equation}
where 
$\pr \hat{\X}|_{\Esp} = \pr \X|_{\Esp} -\tau D_t - \xi D_r$.

An infinitesimal point symmetry is \emph{trivial} 
if its action on the solution space $\Esp$ is trivial,
$\hat{\X}_{\text{triv}}u = 0$ 
for all solutions $u(t,r)$.
This occurs iff $P|_{\Esp}=0$.
The corresponding generator
\eqref{point_symm}
of a trivial symmetry is thus given by
$ \X_{\text{triv}}|_{\Esp}=\tau\partial_t +
\xi\partial_r +
(\tau u_t+\xi u_r)\partial_u$,
which has the prolongation
$\pr\X_{\text{triv}}|_{\Esp} = \tau D_t + \xi D_r$.
Conversely,
any generator of this form on the solution space $\Esp$
determines a trivial symmetry.

The set of all non-trivial infinitesimal point symmetries forms a Lie algebra
(with respect to commutation of the generators). 

The symmetry determining equation \eqref{sym_det_eqn}
splits into a linear overdetermined system of equations,
after $u_t$ and its differential consequences are eliminated through 
the equation \eqref{nonlin_diffus_eqn_1D}--\eqref{diffus_rel_1D}. 
This yields the determining system
\begin{subequations}\label{symm_det_sys}
\begin{align}
& \tau_r = 0 \\
& \tau_u = 0\\
& (u_rh''+2h')(u_r \xi_u+\xi_r)-h''(u_r\eta_u+\eta_r) -\tau_t h' = 0\\
& \begin{aligned}
& rh'\big(u_r^3\xi_{uu}+u_r^2(2\xi_{ur}-\eta_{uu})+u_r(\xi_{rr}-2\eta_{ur})-\eta_{rr}\big)
+\big(m(u_rh'-h)-rf\big)(u_r\xi_u-\eta_u) \\
& \qquad - (mh+rf)\tau_t +m\big(h'(u_r\xi_r-\eta_r)+r^{-1}h\xi\big) 
-r(u_r\xi_t-\eta_t+f'\eta) = 0
\end{aligned}
\end{align}
\end{subequations}
for $\eta(t,r,u)$,  $\tau(t,r,u)$,  $\xi(t,r,u)$, $h(u_r)$, $f(u)$, $m$. 
Note, from equation \eqref{nonlin_diffus_eqn_1D}, 
we have the conditions $h''\neq 0$ and $f\neq 0$, 
which respectively imply the diffusivity is non-constant and the source/sink is non-zero. 
We will also impose the conditions $(u_r h)'\neq0$ and $((u_r h)'/h)'=f'\neq0$
which exclude all cases in which the hodograph transformation exists, 
as shown in Proposition~\ref{hodograph}. 

The determining system \eqref{symm_det_sys} can be split with respect to $u_r$
in a similar way to the splitting of the multiplier determining system \eqref{mult_det_sys}.
To derive the split system and obtain its complete set of solutions, 
we have used the steps outlined in the appendix. 

The components $(\eta(t,r,u),\tau(t,r,u),\xi(t,r,u))$ of 
a complete set of point symmetry generators,
excluding the hodograph cases, 
are listed in table \ref{table_pointsymm}. 
The table is arranged by, 
firstly, the dimension of the maximal symmetry Lie algebras,
and secondly, the generality of $h(u_r)$, $f(u)$, $m$,
where the double lines separate cases of different dimensionality. 

\begin{footnotesize}
\begin{table}[htb]
\centering
\caption{Generators of point symmetries (in non-hodograph cases)}
\label{table_pointsymm}
\begin{tabular}{c c c | c c r}
\hline
$\tau$ 
& $\xi$ 
& $\eta$
& $h(u_r)$
& $f(u)$ 
& $m$  
\\
\hline
\hline
$1$
& $0$
& $0$
&
&
& 
\\
\hline
\hline
$0$
& $1$
& $0$
&
&
& $0$
\\
\hline
$0$
& $0$
& $e^{kt}$
& 
& $b+ku$
&
\\
\hline
$2t$
& $r$
& $a+u$
&
& $k/(a+u)$
& 
\\
\hline
$\frac{1}{c}e^{c(1-p)t}$
& $0$
& $e^{c(1-p)t}(a+u)$
& $-\kappa u_r^{p}$
& $k(a+u)^{p}$
&
\\
& 
& 
& $p \neq -1,0,1$
& \quad $+c(a+u)$
&
\\
& 
& 
& 
& $c \neq 0$
&
\\
\hline
$(p+1)(1-q)t$
& $(p-q)r$
& $(p+1)(a+u)$
& $-\kappa u_r^{p}$
& $k(a+u)^{q}$ 
& 
\\
& 
& 
& $p\neq -1,0,1$
& $q\neq -1,0$
&
\\
\hline
$(p+1)qt$
& $qr$
& $-(p+1)$
& $-\kappa u_r^{p}$
& $k e^{qu}$ 
& 
\\
& 
& 
& $p \neq -1,0,1$
& $q\neq 0$
&
\\
\hline
$2t$
& $(1-\frac{b}{2\beta}r)r$
& $a+u$
& $\beta-\kappa/u_r$
& $b+k/(a+u)$
& $-2$
\\
& 
& 
& $\beta \neq 0$
& $k\neq 0$, $b\neq 0$
&
\\
\hline
\hline
$2t$
& $r$
& $bt+u$
& 
& $b$
& 
\\
\hline
$-kt^2$
& $0$
& $1+2kt(a+u)$
& $-\kappa u_r^{2}$
& $k(a+u)^2$
&
\\
\hline
$-\frac{1}{2b}e^{2bt}$
& $0$
& $e^{2bt} (a+\frac{b}{k} +u)$
& $-\kappa u_r^{2}$
& $k(a+u)^2-b^2/k$
& 
\\
& 
& 
&
& $k\neq 0$, $b\neq 0$
&
\\
\hline
$\frac{1}{2b}\cos(2bt)$
& $0$
& $\frac{b}{k}\cos(2bt)$
& $-\kappa u_r^{2}$
& $k(a+u)^2+b^2/k$
& 
\\
& 
& \quad $+\sin(2bt)(a+u)$
& 
& $k\neq 0$, $b\neq 0$
& 
\\
$\frac{1}{2b}\sin(2bt)$
& $0$
& $\frac{b}{k}\sin(2bt)$
&
&
& 
\\
& 
& \quad $-\cos(2bt)(a+u)$
& 
& 
& 
\\
\hline
$0$
& $-kr^2$
& $2\beta$
& $\beta-\kappa/u_r$
& $b+ku$
& $-2$
\\
& 
& 
& $\beta \neq 0$
& $k \neq 0$
&
\\
\hline
\hline
$(p+1)t$
& $r$
& $(p+1)bt-\alpha r$
& $-\kappa (\alpha+u_r)^{p}$
& $b$
& 
\\
& 
& 
& $p \neq -1,0,1$
& 
&
\\
\hline
$pt$
& $0$
& $bpt-r$
& $\kappa e^{pu_r}$ 
& $b$
& 
\\
& 
& 
& $p \neq 0$
& 
&
\\
\hline
$\alpha e^{-kt}$
& $-e^{-kt} (b+ku)$
& $\alpha e^{-kt}(b+ku)$
& $-\kappa u_r^2/(\alpha + u_r)^{2}$
& $b+ku$
& $0$
\\
& 
& 
& $\alpha \neq 0$
& $k \neq 0$
&
\\
\hline
$e^{kt}$
& $0$
& $k e^{kt}(\alpha r+u)$
& $\kappa \ln|\alpha + u_r|$
& $b+ku$
& $0$
\\
& 
& 
& 
& $k \neq 0$
&
\\
\hline
\hline
$((p+1)\beta$
& $bt-u$
& $bp(\beta-\alpha)t+\alpha\beta r$
& $-\kappa\big(\frac{\beta+u_r}{\alpha+u_r}\big)^p $
& $b$
& $0$
\\
\quad$-(p-1)\alpha)t$
&
& \quad $+(\alpha+\beta)u$
& $p\neq 0$, $\alpha\neq\beta$
& 
& 
\\
\hline
$(2\alpha-p\beta)t$
& $bt-u$
& $(\alpha^2+\beta^2)r-b\beta pt$
& $\kappa e^{p\arctan((\alpha+u_r)/\beta)}$
& $b$
& $0$
\\
&
& \quad $+2\alpha u$
& $p\neq 0$, $\beta\neq 0$
& 
& 
\\
\hline
$(2\alpha+p)t$
& $bt-u$
& $bpt+\alpha(\alpha r+2u)$
& $\kappa e^{p/(\alpha+u_r)}$
& $b$
& $0$
\\
&
& 
& $p\neq 0$
& 
& 
\\
\hline
$0$
& $r^2$
& $2\beta t-\alpha r^2$
& $\beta-\kappa/(\alpha+u_r)$
& $b$
& $-2$
\\
$2\beta t^2$
& $r^2(\alpha r- bt)$ 
& $\alpha r^2(bt-\alpha r)$ 
& $\beta\neq 0$
& 
& 
\\
& \quad$+r(2\beta t +ru)$
& \quad$+(2\beta t -\alpha r^2)u$
& 
& 
& 
\\
\hline
\end{tabular}
\end{table}
\end{footnotesize}

From the generators shown in the table~\ref{table_pointsymm}, 
we have the following classification result. 

\begin{thm}\label{point_symm_thm}
(i) 
The point symmetries 
admitted by the $1$-dimensional nonlinear diffusion-reaction equation \eqref{nonlin_diffus_eqn_1D}--\eqref{diffus_rel_1D} 
for arbitrary diffusivity $h'(u_r)$ 
are generated by the transformations:
\begin{subequations}\label{point_symm1}
\begin{flalign}
& f(u) \text{ arbitrary}&
\\
& \X_{1} = \partial_t&
\\
& (\tilde{t},\tilde{r},\tilde{u})_1 
= (t+\epsilon,r,u)
\quad \text{time-translation}&
\end{flalign}
\end{subequations}
\begin{subequations}\label{point_symm2}
\begin{flalign}
& f(u) \text{ arbitrary}, m=0&
\\
& \X_{2} = \partial_r&
\\
& (\tilde{t},\tilde{r},\tilde{u})_2 
=(t,r+\epsilon,u)
\quad \text{space-translation}&
\end{flalign}
\end{subequations}
\begin{subequations}\label{point_symm3}
\begin{flalign}
& f(u)=b+ku&
\\
& \X_{3} = \exp(kt) \partial_u&
\\
& (\tilde{t},\tilde{r},\tilde{u})_3 
=(t,r,u+\epsilon \exp(kt))
\quad \text{time-dependent shift}&
\end{flalign}
\end{subequations}
\begin{subequations}\label{point_symm4}
\begin{flalign}
& f(u)=k/(a+u)&
\\
& \X_{4} = 2t \partial_t + r \partial_r + (a+u) \partial_u&
\\
& (\tilde{t},\tilde{r},\tilde{u})_4 
=\big(e^{2 \epsilon}t, e^{\epsilon}r, e^{\epsilon}(a+u)-a \big)
\quad \text{scaling and shift}&
\end{flalign}
\end{subequations}
\begin{subequations}\label{point_symm5}
\begin{flalign}
& f(u)=b&
\\
& \X_{5} = 2t \partial_t + r \partial_r + (bt+u) \partial_u&
\\
& (\tilde{t},\tilde{r},\tilde{u})_5 
=\big(e^{2\epsilon}t, e^{\epsilon}r, e^{\epsilon}((e^{\epsilon}-1)bt+u)\big)
\quad \text{scaling and time-dependent shift}&
\end{flalign}
\end{subequations}
\newline
(ii) 
The $1$-dimensional nonlinear diffusion-reaction equation \eqref{nonlin_diffus_eqn_1D}--\eqref{diffus_rel_1D} 
admits additional point symmetries only for diffusivities $h'(u_r)$ given by 
\begin{flalign}
& (a)\quad 
h(u_r)=\beta-\kappa(\alpha + u_r)^{p}
\label{case-a}\\
& (b)\quad 
h(u_r)=-\kappa\Big(\frac{\beta+u_r}{\alpha+u_r}\Big)^p
\label{case-b}\\
& (c)\quad 
h(u_r)=\kappa\exp(p\arctan((\alpha+u_r)/\beta)) 
\label{case-c}\\
& (d)\quad 
h(u_r)=\kappa\exp(p/(\alpha+u_r))
\label{case-d}\\
& (e)\quad 
h(u_r)=\kappa\exp(pu_r) 
\label{case-3}\\
& (f)\quad 
h(u_r)=\kappa\ln|\alpha + u_r|
\label{case-f}
\end{flalign}
The transformations generating these symmetries in each (non-hodograph) case 
are given by:
\begin{enumerate}[(a)]
\item
\begin{subequations}\label{point_symm6}
\begin{flalign}
&\alpha=0, \beta=0, p \neq -1,0,1;
f(u)=k(a+u)^{p}+c(a+u), 
c \neq 0&
\\
& \X_{6} = c^{-1}\exp(c(1-p)t) \partial_t + \exp(c(1-p)t)(a+u) \partial_u&
\\
&(\tilde{t},\tilde{r},\tilde{u})_{6} 
=\big(c^{-1}(p-1)^{-1}\ln|\exp(c(p-1)t)+\epsilon|, r, (1+\epsilon \exp(c(1-p)t))^{1/(p-1)} (a+u)-a\big)&
\\
\nonumber
&\quad \text{dilation and shift}&
\end{flalign}
\end{subequations}
\begin{subequations}\label{point_symm7}
\begin{flalign}
&\alpha=0, \beta=0, p \neq -1,0,1; 
f(u)=k(a+u)^{q}, 
q\neq -1,0&
\\
& \X_{7} = (p+1)(1-q)t \partial_t + (p-q)r \partial_r +(p+1)(a+u) \partial_u&
\\
&(\tilde{t},\tilde{r},\tilde{u})_{7} 
=\big(e^{(p+1)(1-q) \epsilon}t, e^{(p-q)\epsilon}r, e^{(p+1)\epsilon}(a+u)-a\big)
\quad \text{scaling and shift}&
\end{flalign}
\end{subequations}
\begin{subequations}\label{point_symm8}
\begin{flalign}
&\alpha=0, \beta=0, p \neq -1,0,1; 
f(u)=k e^{qu}, 
q\neq 0&
\\
& \X_{8} = (p+1)qt \partial_t + qr \partial_r -(p+1) \partial_u&
\\
&(\tilde{t},\tilde{r},\tilde{u})_{8} 
=(e^{(p+1)q\epsilon}t,  e^{q\epsilon}r, u-(p+1)\epsilon)
\quad \text{scaling and shift}&
\end{flalign}
\end{subequations}
\begin{subequations}\label{point_symm9}
\begin{flalign}
&\alpha=0, \beta\neq 0, p=-1; 
f(u)=b+k/(a+u), k\neq 0, b\neq 0; 
m=-2&
\\
& \X_{9} = 2t \partial_t + (1-\tfrac{1}{2}b\beta^{-1}r)r \partial_r + (a+u) \partial_u&
\\
&(\tilde{t},\tilde{r},\tilde{u})_{9} =
\big(e^{2\epsilon}t,  
2\beta(e^{-\epsilon}(b-2\beta r^{-1})-b)^{-1}, 
e^{\epsilon}(a+u)-a\big)
\quad \text{dilation and shift}&
\end{flalign}
\end{subequations}
\begin{subequations}\label{point_symm10}
\begin{flalign}
&\alpha=0, \beta=0, p=2; 
f(u)=k(a+u)^2&
\\
& \X_{10} = -kt^2 \partial_t + (1+2kt(a+u)) \partial_u&
\\
& (\tilde{t},\tilde{r},\tilde{u})_{10} =
\big((1 + k\epsilon t)^{-1}t, r, 
(1 + k\epsilon t)( (1 + k\epsilon t)(a+u)+\epsilon)-a \big)
\quad \text{dilation and shift}&
\end{flalign}
\end{subequations}
\begin{subequations}\label{point_symm11}
\begin{flalign}
& \alpha=0, \beta=0, p=2; 
f(u)=k(a+u)^2-b^2/k, 
k\neq 0, b\neq 0
&\\
& \X_{11} = -\tfrac{1}{2}b^{-1}\exp(2bt) \partial_t + \exp(2bt) (a+bk^{-1} +u) \partial_u&\\
& (\tilde{t},\tilde{r},\tilde{u})_{11} =
\big(
{-}\tfrac{1}{2}b^{-1} \ln|\exp(-2bt)+\epsilon|, 
r,  
\exp(2bt)\epsilon(a+bk^{-1}+u)+u\big)
\\
\nonumber
&\quad \text{exponential dilation and shift}&
\end{flalign}
\end{subequations}
\begin{subequations}\label{point_symm12-13}
\begin{flalign}
&\alpha=0, \beta=0, p=2; 
f(u)=k(a+u)^2+b^2/k, 
k\neq 0, b\neq 0&
\\
& \X_{12} = \tfrac{1}{2}b^{-1}\cos(2bt) \partial_t + (bk^{-1}\cos(2bt)+\sin(2bt)(a+u)) \partial_u&
\\
& \begin{aligned}
(\tilde{t},\tilde{r},\tilde{u})_{12} =
&\Big(b^{-1}\arctan\big((\sec(2bt)+\cosh(\epsilon))^{-1}(\tan(2bt)+\sinh(\epsilon))\big), r, 
\\
&\qquad
bk^{-1}\cos(2bt)\sinh(\epsilon)+(\cosh(\epsilon)+\sin(2bt)\sinh(\epsilon))(a+u)-a\Big)
\end{aligned}&
\\
\nonumber
&\quad \text{oscillatory dilation and shift}
\\
& \X_{13} = \tfrac{1}{2}b^{-1}\sin(2bt) \partial_t + (bk^{-1}\sin(2bt)-\cos(2bt)(a+u)) \partial_u&
\\
&\begin{aligned}
(\tilde{t},\tilde{r},\tilde{u})_{13} =
&\Big(\tfrac{1}{b}\arctan\big((\csc(2bt)+\cosh(\epsilon))^{-1}(-\cot(2bt)+\sinh(\epsilon))\big), r, 
\\
&\qquad
bk^{-1}\sin(2bt)\sinh(\epsilon)+(\cosh(\epsilon)-\cos(2bt)\sinh(\epsilon))(a+u)-a\Big)
\end{aligned}&
\\
\nonumber
&\quad \text{oscillatory dilation and shift}&
\end{flalign}
\end{subequations}
\begin{subequations}\label{point_symm14}
\begin{flalign}
&\alpha=0, \beta\neq 0, p=-1;
f(u)=b+ku, k\neq 0; 
m=-2&
\\
& \X_{14} = -kr^2 \partial_r + 2\beta \partial_u&
\\
& (\tilde{t},\tilde{r},\tilde{u})_{14} =
\big(t, r(1+k\epsilon r)^{-1}, 2\beta\epsilon+u \big)
\quad \text{dilation and shift}&
\end{flalign}
\end{subequations}
\begin{subequations}\label{point_symm15}
\begin{flalign}
&\beta=0, p\neq -1,0,1;
f(u)=b &
\\
& \X_{15} = (p+1)t \partial_t + r \partial_r + ((p+1)bt-\alpha r) \partial_u&
\\
&(\tilde{t},\tilde{r},\tilde{u})_{15} =
\big(e^{(p+1)\epsilon}t,  e^{\epsilon}r, (e^{(p+1)\epsilon}-1)bt-(e^{\epsilon}-1)\alpha r+u\big)&
\\
\nonumber
&\quad \text{scaling and time- $\&$ space- dependent shift}&
\end{flalign}
\end{subequations}
\begin{subequations}\label{point_symm16-17}
\begin{flalign}
& \beta\neq 0, p=-1; 
f(u)=b; 
m=-2&
\\
& \X_{16} = r^2 \partial_r + (2\beta t-\alpha r^2) \partial_u&
\\
& (\tilde{t},\tilde{r},\tilde{u})_{16} =
\big(t,  r(1-\epsilon r)^{-1}, 
\epsilon(2\beta t-\alpha r^2(1-\epsilon r)^{-1}) +u \big)&
\\
\nonumber
&\quad \text{dilation and time- $\&$ space- dependent shift}&
\\
& \X_{17} = 2\beta t^2 \partial_t + (r^2(\alpha r- bt)+r(2\beta t +ru)) \partial_r + (\alpha r^2 (bt-\alpha r)+(2\beta t -\alpha r^2)u) \partial_u&
\\
&\begin{aligned}	
(\tilde{t},\tilde{r},\tilde{u})_{17} =
& \Big(
t(1-2\beta \epsilon t)^{-1},  
r\big(1-\epsilon(2\beta t+r(\alpha r-bt+u))\big)^{-1}, 
\\
& \qquad
(1-2\beta \epsilon t)^{-1}\big(\alpha\epsilon r^2(\alpha r-bt+u)(1-\epsilon(2\beta t+r(\alpha r-bt+u)))^{-1}+u\big)
\Big)
\end{aligned}&
\\
\nonumber
&\quad \text{dilation and time- $\&$ space- dependent shift}&
\end{flalign}
\end{subequations}
\item
\begin{subequations}\label{point_symm18-19}
\begin{flalign}
& \alpha \neq 0, \beta=0, p=2; 
f(u)=b+ku, k \neq 0; 
m=0&
\\
& \X_{18} = \alpha \exp(-kt) \partial_t -\exp(-kt) (b+ku)\partial_r + \alpha \exp(-kt)(b+ku) \partial_u&
\\
& (\tilde{t},\tilde{r},\tilde{u})_{18} =
\big(\tfrac{1}{k}\ln|\exp(kt)+\alpha k\epsilon|,  
-\epsilon \exp(-kt)(b+ku)+r, 
\alpha \epsilon \exp(-kt)(b+ku) +u\big)&
\\
\nonumber
&\quad \text{exponential dilation and shift}&
\end{flalign}
\end{subequations}
\begin{subequations}
\begin{flalign}
& \alpha\neq\beta, p \neq 0;
f(u)=b; 
m=0&
\\
& \X_{19} = ((p+1)\beta-(p-1)\alpha)t\partial_t +(bt-u)\partial_r + (bp(\beta-\alpha)t+\alpha\beta r+(\alpha+\beta)u)\partial_u&
\\
&\begin{aligned} 
(\tilde{t},\tilde{r},\tilde{u})_{19} =
& \Big(e^{((p+1)\beta-(p-1)\alpha)\epsilon}t, 
\tfrac{1}{\beta-\alpha}\big((e^{\beta\epsilon}-e^{\alpha\epsilon})(bt-u)+(\beta e^{\alpha\epsilon}-\alpha e^{\beta\epsilon})r\big), 
\\
& \qquad be^{((p+1)\beta-(p-1)\alpha)\epsilon}t+\tfrac{1}{\beta-\alpha}\big(\alpha\beta(e^{\beta\epsilon}-e^{\alpha\epsilon})r+(\alpha e^{\alpha\epsilon}-\beta e^{\beta\epsilon})(bt-u)\big)
\Big)
\end{aligned}&
\end{flalign}
\end{subequations}
\item
\begin{subequations}\label{point_symm20}
\begin{flalign}
& \beta \neq 0, p \neq 0;
f(u)=b; 
m=0&
\\
& \X_{20} = (2\alpha-p\beta)t \partial_t +(bt-u)\partial_r + ((\alpha^2+\beta^2)r-b\beta pt+2\alpha u) \partial_u&
\\
& \begin{aligned} (\tilde{t},\tilde{r},\tilde{u})_{20} =& \Big( e^{(\beta p-2\alpha)\epsilon}t, 
e^{-\alpha\epsilon}\big(\cos(\beta\epsilon)r+\tfrac{1}{\beta}\sin(\beta\epsilon)(b t+\alpha r-u)\big), 
\\
&\qquad 
be^{(\beta p-2\alpha)\epsilon}t
+e^{-\alpha\epsilon}\big(\tfrac{1}{\beta}\sin(\beta\epsilon)(\alpha(bt-u)-(\alpha^2+\beta^2)r)-\cos(\beta\epsilon)(bt-u)\big)\Big) 
\end{aligned} &
\end{flalign}
\end{subequations}
\item
\begin{subequations}\label{point_symm21}
\begin{flalign}
& p \neq 0; 
f(u)=b; 
m=0&
\\
& \X_{21} = (2\alpha+p)t \partial_t + (bt-u) \partial_r + (bpt+\alpha(\alpha r+2u)) \partial_u&
\\
& (\tilde{t},\tilde{r},\tilde{u})_{21} =
\Big(e^{(p+2\alpha)\epsilon}t,  
e^{\alpha\epsilon}(\epsilon(bt-\alpha r-u)+r), 
be^{(p+2\alpha)\epsilon}t-e^{\alpha\epsilon}\big(\alpha r+(1+\alpha\epsilon)(bt-\alpha r-u)\big)\Big)&
\end{flalign}
\end{subequations}
\item
\begin{subequations}\label{point_symm22}
\begin{flalign}
& p \neq 0; 
f(u)=b &
\\
& \X_{22} = pt \partial_t + (bpt-r) \partial_u&
\\
& (\tilde{t},\tilde{r},\tilde{u})_{22} =
\big(e^{p\epsilon}t,  r, bt(e^{p\epsilon}-1)-\epsilon r+u\big)
\quad \text{dilation and time- $\&$ space- dependent shift}&
\end{flalign}
\end{subequations}
\item
\begin{subequations}\label{point_symm23}
\begin{flalign}
& f(u)=b+ku, 
k \neq 0; 
m=0&
\\
& \X_{23} = \exp(kt)\partial_t + k\exp(kt)(\alpha r+u) \partial_u&
\\
& (\tilde{t},\tilde{r},\tilde{u})_{23} =
\big({-}k^{-1}\ln|\exp(-kt)-\epsilon|,  
r, (1-\epsilon\exp(kt))^{-1}(\alpha r\epsilon\exp(kt)+u)\big)&
\\
\nonumber
&\quad \text{exponential dilation and shift}&
\end{flalign}
\end{subequations}
\end{enumerate}
\end{thm}

We now state the classification of the point-symmetry Lie algebras 
for the class of $1$-dimensional nonlinear diffusion-reaction equation \eqref{nonlin_diffus_eqn_1D}--\eqref{diffus_rel_1D}. 
In particular, 
a basis of generators for each maximal Lie algebra is exhibited for each non-hodograph case. 

\begin{thm}\label{symm_groups}
The maximal point symmetry groups for the class of $1$-dimensional nonlinear diffusion-reaction equation \eqref{nonlin_diffus_eqn_1D}--\eqref{diffus_rel_1D} 
(in all non-hodograph cases)
are generated by:
\begin{enumerate}[(i)]
\item \quad
\newline
arbitrary $h(u_r)$, $f(u)$
\newline
$\X_{1}$, arbitrary $m$; 
$\X_{2}$, $m=0$

\item \quad
\newline
arbitrary $h(u_r)$, $f(u)=b+ku$; $k\neq 0$
\newline
$\X_{1}$, $\X_{3}$, arbitrary $m$; 
$\X_{2}$, $m=0$
\begin{equation}\label{2dim_1}
\left[\X_{1}, \X_{3}\right]= k \X_{3}
\end{equation}
arbitrary $h(u_r)$, $f(u)=k/(a+u)$
\newline
$\X_{1}$, $\X_{4}$, arbitrary $m$; 
$\X_{2}$, $m=0$
\begin{subequations}\label{2dim_2}
\begin{align}
&\left[\X_{1}, \X_{4}\right] = 2 \X_{1};\\
&\left[\X_{2}, \X_{4}\right] = \X_{2} 
\end{align}
\end{subequations}
$h(u_r)=-\kappa u_r^{p}$, $f(u)=k(a+u)^{p}+c(a+u)$; 
$p\neq-1,0,1$; $c\neq 0$
\newline
$\X_{1}$, $\X_{6}$, arbitrary $m$; 
$\X_{2}$, $m=0$
\begin{equation}\label{2dim_3}
\left[\X_{1}, \X_{6}\right]=c(1-p) \X_{6}
\end{equation}
$h(u_r)=-\kappa u_r^{p}$, $f(u)=k(a+u)^{q}$; 
$p \neq -1,0,1$; $q\neq -1,0$
\newline
$\X_{1}$, $\X_{7}$, arbitrary $m$; 
$\X_{2}$, $m=0$
\begin{subequations}\label{2dim_4}
\begin{align}
&\left[\X_{1}, \X_{7}\right]=(p+1)(1-q) \X_{1};\\
&\left[\X_{2}, \X_{7}\right]=(p-q) \X_{2}
\end{align}
\end{subequations}
$h(u_r)=-\kappa u_r^{p}$, $f(u)=k\exp(qu)$; 
$p \neq -1,0,1$; $q\neq 0$
\newline
$\X_{1}$, $\X_{8}$, arbitrary $m$; 
$\X_{2}$, $m=0$
\begin{subequations}\label{2dim_5}
\begin{align}
&\left[\X_{1}, \X_{8}\right]=(p+1)q\X_{1};\\
&\left[\X_{2}, \X_{8}\right]=q \X_{2}
\end{align}
\end{subequations}
$h(u_r)=\beta - \kappa/u_r$, $f(u)=b+k/(a+u)$, $m=-2$; 
$\beta\neq 0$; $k\neq 0$; $b\neq 0$
\newline
$\X_{1}$, $\X_{9}$
\begin{equation}\label{2dim_6}
\left[\X_{1}, \X_{9}\right]=2\X_{1}
\end{equation}

\item \quad
\newline
arbitrary $h(u_r)$, $f(u)=b$
\newline
$\X_{1}$, $\X_{3}|_{k=0}$, $\X_{5}$, arbitrary $m$; $\X_{2}$, $m=0$
\begin{subequations}\label{3dim_1}
\begin{align}
&\left[\X_{1}, \X_{5}\right] = 2 \X_{1} + b \X_{3},
\quad
\left[\X_{3}, \X_{5}\right] = \X_{3};\\
&\left[\X_{2}, \X_{5}\right] = \X_{2}
\end{align}
\end{subequations}
$h(u_r)=-\kappa u_r^{2}$, $f(u)=k(a+u)^2$
\newline
$\X_{1}$, $\X_{7}|_{p=q=2}$, $\X_{10}$, arbitrary $m$; $\X_{2}$, $m=0$
\begin{equation}\label{3dim_2}
\left[\X_{1}, \X_{7}\right]=-3 \X_{1},
\quad
\left[\X_{1}, \X_{10}\right]=\tfrac{2}{3}k \X_{7},
\quad
\left[\X_{7}, \X_{10}\right]=-3 \X_{10}
\end{equation}
$h(u_r)=-\kappa u_r^{2}$, $f(u)=k(a+u)^2-b^2/k$; 
$k\neq 0$; $b\neq 0$
\newline
$\X_{1}$, $\X_{6}|_{p=2,a=a-b/k,c=2b}$, $\X_{11}$, arbitrary $m$; 
$\X_{2}$, $m=0$
\begin{equation}\label{3dim_3}
\left[\X_{1}, \X_{6}\right]=-2b \X_{6},
\quad
\left[\X_{1}, \X_{11}\right]=2b \X_{11},
\quad
\left[\X_{6}, \X_{11}\right]=-\tfrac{1}{b} \X_{1}
\end{equation}
$h(u_r)=-\kappa u_r^{2}$, $f(u)=k(a+u)^2+b^2/k$; 
$k\neq 0$; $b\neq 0$
\newline
$\X_{1}$, $\X_{12}$, $\X_{13}$, arbitrary $m$; 
$\X_{2}$, $m=0$
\begin{equation}\label{3dim_4}
\left[\X_{1}, \X_{12}\right]=-2b\X_{13},
\quad
\left[\X_{1}, \X_{13}\right]=2b\X_{12},
\quad
\left[\X_{12}, \X_{13}\right]=\tfrac{1}{2b}\X_{1}
\end{equation}
$h(u_r)=\beta - \kappa/u_r$, $f(u)=b+ku$, $m=-2$; 
$\beta\neq 0$; $k\neq 0$
\newline
$\X_{1}$, $\X_{3}$, $\X_{14}$
\begin{equation}\label{3dim_5}
\left[\X_{1}, \X_{3}\right]=k \X_{3}
\end{equation}

\item \quad
\newline
$h(u_r)=-\kappa (\alpha+u_r)^{p}$, $f(u)=b$; 
$p\neq -1,0,1$
\newline
$\X_{1}$, $\X_{3}|_{k=0}$, $\X_{5}$, $\X_{15}$, arbitrary $m$; 
$\X_{2}$, $m=0$
\begin{subequations}\label{4dim_1}
\begin{align}
&\left[\X_{1}, \X_{5}\right]
=2\X_{1}+b\X_{3},
\quad
\left[\X_{1}, \X_{15}\right]=(p+1)(\X_{1}+b\X_{3}),
\quad
\left[\X_{3}, \X_{5}\right]=\X_{3};
\\
&\left[\X_{2}, \X_{5}\right]=\X_{2},
\quad 
\left[\X_{2}, \X_{15}\right]=\X_{2}- \alpha \X_{3}
\end{align}
\end{subequations}
$h(u_r)=\kappa \exp(pu_r)$, $f(u)=b$;
$p \neq 0$
\newline
$\X_{1}$, $\X_{3}|_{k=0}$, $\X_{5}$, $\X_{22}$, arbitrary $m$; 
$\X_{2}$, $m=0$
\begin{subequations}\label{4dim_2}
\begin{align}
&\left[\X_{1}, \X_{5}\right]= 2 \X_{1} + b \X_{3},
\quad
\left[\X_{1}, \X_{22}\right]=p(\X_{1}+b\X_{3}),
\quad
\left[\X_{3}, \X_{5}\right] = \X_{3};\\
&\left[\X_{2}, \X_{5}\right] = \X_{2},
\quad 
\left[\X_{2}, \X_{22}\right]=-\X_{3}
\end{align}
\end{subequations}
$h(u_r)=-\kappa u_r^{p}$, $f(u)=b+ku$; 
$p \neq -1,0,1$
\newline
$\X_{1}$, $\X_{3}$, $\X_{6}|_{k=0,c=k,a=b/k}$, $\X_{7}|_{q=1,a=b/k}$, arbitrary $m$; 
$\X_{2}$, $m=0$
\begin{subequations}\label{4dim_3}
\begin{align}
&\left[\X_{1}, \X_{3}\right] = k\X_{3},
\quad
\left[\X_{1}, \X_{6}\right] = k(1-p)\X_{6},
\quad
\left[\X_{3}, \X_{7}\right] = (p+1)\X_{3};\\
& \left[\X_{2}, \X_{7}\right] = (p-1)\X_{2}
\end{align}
\end{subequations}
$h(u_r)=-\kappa u_r^2/(\alpha + u_r)^{2}$, $f(u)=b+ku$, $m=0$;
$\alpha\neq 0$; $k\neq 0$
\newline
$\X_{1}$, $\X_{2}$, $\X_{3}|_{m=0}$, $\X_{18}$
\begin{equation}\label{4dim_4}
\left[\X_{1}, \X_{3}\right]=k \X_{3},
\quad
\left[\X_{1}, \X_{18}\right]=-k \X_{18},
\quad
\left[\X_{3}, \X_{18}\right]=-k \X_{2}
\end{equation}	
$h(u_r)=\kappa \ln|\alpha + u_r|$, $f(u)=b+ku$, $m=0$; 
$k \neq 0$
\newline
$\X_{1}$, $\X_{2}$, $\X_{3}|_{m=0}$, $\X_{23}$
\begin{equation}\label{4dim_5}
\left[\X_{1}, \X_{3}\right]=k \X_{3},
\quad
\left[\X_{1}, \X_{23}\right]=k \X_{23},
\quad
\left[\X_{2}, \X_{23}\right]=\alpha k \X_{3}
\end{equation}

\item\quad
\newline
$h(u_r)=-\kappa\Big(\dfrac{\beta+u_r}{\alpha+u_r}\Big)^p$, $f(u)=b$, $m=0$; 
$p\neq 0$; $\alpha\neq\beta$
\newline
$\X_{1}$, $\X_{2}$, $\X_{3}|_{k=0,m=0}$, $\X_{5}$, $\X_{19}$
\begin{subequations}\label{5dim_1}
\begin{align}
&\left[\X_{1}, \X_{5}\right]= 2\X_{1}+b\X_{3},
\quad
\left[\X_{1}, \X_{19}\right]=
((p+1)\beta-(p-1)\alpha+)\X_{1}+b\X_{2}+bp(\beta-\alpha)\X_{3},\\
&\left[\X_{2}, \X_{5}\right]= \X_{2},
\quad
\left[\X_{2}, \X_{19}\right]=
\alpha\beta\X_{3},
\quad
\left[\X_{3}, \X_{5}\right]=
\X_{3},
\quad
\left[\X_{3}, \X_{19}\right]=
(\alpha+\beta)\X_{3}-\X_{2}
\end{align}
\end{subequations}
$h(u_r)= \kappa \exp(p\arctan((\alpha+u_r)/\beta))$, $f(u)=b$,  $m=0$; 
$p\neq 0$; $\beta \neq 0$
\newline
$\X_{1}$, $\X_{2}$, $\X_{3}|_{k=0,m=0}$, $\X_{5}$, $\X_{20}$
\begin{subequations}\label{5dim_2}
\begin{align}
&\left[\X_{1}, \X_{5}\right]= 2\X_{1}+b\X_{3},
\quad
\left[\X_{1}, \X_{20}\right]=
(p\beta-2\alpha)\X_{1}-b\X_{2}+bp\beta\X_{3},
\quad
\left[\X_{2}, \X_{5}\right]= \X_{2},\\
&\left[\X_{2}, \X_{20}\right]=
-(\alpha^2+\beta^2)\X_{3},
\quad
\left[\X_{3}, \X_{5}\right]=
\X_{3},
\quad
\left[\X_{3}, \X_{20}\right]=
\X_{2}-2\alpha\X_{3}
\end{align}
\end{subequations}
$h(u_r)=\kappa\exp(p/(\alpha+u_r))$, $f(u)=b$, $m=0$; 
$p \neq 0$ 
\newline
$\X_{1}$, $\X_{2}$, $\X_{3}|_{k=0,m=0}$, $\X_{5}$, $\X_{21}$
\begin{subequations}\label{5dim_3}
\begin{align}
&\left[\X_{1}, \X_{5}\right]= 2\X_{1}+b\X_{3},
\quad
\left[\X_{1}, \X_{21}\right]=
(p+2\alpha)\X_{1}+b\X_{2}+bp\X_{3},
\quad
\left[\X_{2}, \X_{5}\right]= \X_{2},\\
&\left[\X_{2}, \X_{21}\right]=
\alpha^2\X_{3},
\quad
\left[\X_{3}, \X_{5}\right]=
\X_{3},
\quad
\left[\X_{3}, \X_{21}\right]=
2\alpha\X_{3}-\X_{2}
\end{align}
\end{subequations}
$h(u_r)=\beta - \kappa/(\alpha+u_r)$, $f(u)=b$, $m=-2$; 
$\beta\neq 0$
\newline
$\X_{1}$, $\X_{3}|_{k=0}$, $\X_{5}$, $\X_{16}$, $\X_{17}$
\begin{subequations}\label{5dim_4}
\begin{align}
&\left[\X_{1}, \X_{5}\right] = 2 \X_{1} + b \X_{3},
\quad
\left[\X_{1}, \X_{16}\right]=2\beta \X_{3},
\quad
\left[\X_{1}, \X_{17}\right]=2\beta\X_{5}-b \X_{16},\\
& \left[\X_{3}, \X_{5}\right] = \X_{3},
\quad
\left[\X_{3}, \X_{17}\right]=\X_{16},
\quad
\left[\X_{5}, \X_{16}\right]=\X_{16},
\quad
\left[\X_{5}, \X_{17}\right]=2 \X_{17}
\end{align}
\end{subequations}
\end{enumerate}
\end{thm}

In the literature, 
a classification of point symmetries is known for the case $m=0$ and $f'(u)\neq0$ 
\cite{CheKinKov}, 
which describes a reaction-diffusion equation with a gradient diffusivity in one spatial dimension, $u_t = h'(u_r)u_{rr}+f(u)$. 
We remark that this classification coincides with the classification 
presented in Theorems~\ref{point_symm_thm} and~\ref{symm_groups} 
in the special case $m=0$, 
except that the point symmetry $\X_{18}$ appears to be missing in \Ref{CheKinKov}. 

We also remark that the case $m=0$ and $f(u)= b$ 
as well as the case $m=0$, $f(u)= b+ku$, and $h(u_r)=-\kappa u_r{}^{p}$ 
can be mapped by a point transformation $(t,u)\to (\tilde{t},\tilde{u})$
into a nonlinear filtration equation in one spatial dimension, 
$\tilde{u}_{\tilde{t}} = h'(\tilde{u}_r)\tilde{u}_{rr}$.
The results in Theorems~\ref{point_symm_thm} and~\ref{symm_groups} 
agree with the classification of point symmetries 
known \cite{IbrAkhGaz} for this special class of equations.

\subsection{Contact symmetries}

As the $1$-dimensional nonlinear diffusion-reaction equation \eqref{nonlin_diffus_eqn_1D}--\eqref{diffus_rel_1D} 
involves only a single dependent variable $u$, 
its Lie symmetry group comprises point symmetries and contact symmetries
\cite{BA}. 

A {\em contact symmetry} extends the definition of invariance \eqref{sym_det_eqn}
by allowing the symmetry transformations to depend 
essentially on first-order derivatives of $u$,
as given by an infinitesimal generator with the characteristic form 
\begin{equation}\label{contact_symm_char}
\hat\X=P(t,r,u,u_t,u_r)\parder{u} .
\end{equation}
The corresponding infinitesimal transformations on $(t,x,u,u_t,u_x)$ are given by 
\begin{equation}\label{contact_symm}
\X=\tau\parder{t} +\xi\parder{r} +\eta\parder{u} +\eta^t\parder{u_t} +\eta^r\parder{u
_r}
\end{equation}
where
\begin{equation}\label{contact_symm_form}
\tau= -P_{u_t},
\quad
\xi= -P_{u_r},
\quad
\eta= P-u_t P_{u_t} -u_r P_{u_r},
\quad
\eta^t= P_{t} + u_t P_u,
\quad
\eta^r= P_{r} + u_r P_u 
\end{equation}
which follows from preservation of the contact condition $du =u_t dt + u_r dr$.
Note that a contact symmetry reduces to a (prolonged) point symmetry 
if and only if $P$ is a linear function of $u_t$ and $u_r$. 

In the symmetry determining equation \eqref{sym_det_eqn}, 
because $u_t$ appears in the contact symmetry generator \eqref{contact_symm_char}, 
we eliminate $u_{rr}$ and its differential consequences 
by use of the nonlinear diffusion-reaction equation \eqref{nonlin_diffus_eqn_1D}--\eqref{diffus_rel_1D}. 
Then the determining equation \eqref{sym_det_eqn} 
can be split into a linear overdetermined system 
for $P(t,r,u,u_t,u_r)$, $h(u_r)$, $f(u)$, $m$. 
Note there is no further splitting of the system. 
Then we apply the steps outlined in the appendix
to obtain the complete set of solutions,
and we remove any solutions in which $P$ is linear in both $u_t$ and $u_r$. 

\begin{thm}
The $1$-dimensional nonlinear diffusion-reaction equation \eqref{nonlin_diffus_eqn_1D}--\eqref{diffus_rel_1D} 
(in all non-hodograph cases) 
does not admit any contact symmetries 
other than prolongations of point symmetries. 
\end{thm}

\section{Low-order conservation laws and Lie symmetries in $n\geq1$ dimensions}
\label{lift_sec}

The solution space $\Esp$ of the $1$-dimensional diffusion-reaction equation \eqref{nonlin_diffus_eqn_1D}--\eqref{diffus_rel_1D} 
coincides with the subspace $\Esp\subset \Esp_+$ of solutions $u(t,r)$ 
of the radial diffusion-reaction equation \eqref{nonlin_diffus_eqn_rad}
without the regularity conditions \eqref{regularity}
iff $h(u_r)$ is an odd function 
or equivalently iff the diffusivity $h'(u_r)$ is an even function.
As a consequence of this relation between the solution spaces, 
if a local conservation law admitted by the $1$-dimensional equation 
when $h(u_r)$ is an odd function 
is restricted to solutions satisfying the regularity conditions \eqref{regularity}, 
then the conservation law is also admitted by the radial equation. 
Similarly, 
if a Lie symmetry admitted by the $1$-dimensional equation 
when $h(u_r)$ is an odd function 
preserves the regularity conditions \eqref{regularity}, 
then the symmetry is admitted by the radial equation. 
Conversely, 
if a local conservation law admitted by the radial equation 
holds without the regularity conditions \eqref{regularity},
then the conservation law can be extended to the $1$-dimensional equation, 
with $h(u_r)=g(|u_r|)u_r$ being an odd function. 
Likewise, if a Lie symmetry admitted by the radial equation 
holds without preserving the regularity conditions \eqref{regularity}, 
then the symmetry can be extended to the $1$-dimensional equation. 

Hence, 
the classification of low-order local conservation laws and Lie symmetries 
shown in Theorems~\ref{cons_laws_thm} and~\ref{point_symm_thm}
for the $1$-dimensional equation directly yields 
a corresponding classification for the radial equation. 

We start with the classification of conservation laws. 

\begin{thm}\label{cons_laws_thm_rad}
The low-order local conservation laws 
\begin{equation}\label{cons_law_rad}
(D_t T +D_r X)|_{\Esp_+}=0
\end{equation}
admitted on the whole solution space $\Esp_+$ of 
the radial generalized $p$-Laplacian diffusion-reaction equation \eqref{nonlin_diffus_eqn_rad} 
are given by:
\begin{subequations}\label{cons_law1_rad}
\begin{flalign}
& 
g(|u_r|) \text{ arbitrary}, 
\quad
f(u) = b+ku &
\\
& 
T_1=r^{n-1} \exp(-kt) u,
\quad
X_1 = - r^{n-1} \exp(-kt) (u_rg(|u_r|) + (b/n)r) & 
\end{flalign}
\end{subequations}
\begin{subequations}\label{cons_law2_rad}
\begin{flalign}
& 
g(|u_r|)= -\kappa/|u_r|^2, 
\quad
f(u) \text{ arbitrary }&
\label{case2_rad}
\\
& 
T_{2,\infty}= r^{n-1} \phi(t,u),
\quad
X_{2,\infty}= -\kappa r^{n-1} \frac{\phi_u(t,u)}{u_r} &
\\
& \phi_t +\kappa\phi_{uu}+f(u)\phi_{u} = 0&
\label{lin_eqn_rad}
\end{flalign}
\end{subequations}
modulo a trivial conserved density and radial flux. 
The corresponding multipliers are given by 
\begin{equation}\label{Q_rad}
Q_1= r^{n-1} u,
\quad
Q_{2,\infty}= r^{n-1} \phi(t,u)
\end{equation}
\end{thm}

We will now discuss the physical meaning of these conservation laws.  

The first conservation law \eqref{cons_law1_rad} 
gives rise to the global continuity equation 
\begin{equation}\label{balance1_rad}
\frac{d}{dt} \int_{0}^{R} e^{-kt} u r^{n-1}dr = 
bR^n/n + e^{-kt} (r^{n-1}g(|\nabla u|) u_r)\Big|_{0}^{R}
\end{equation} 
on a closed radial domain $\Omega=[0,R]\in\Rnum$, with $R>0$. 
This yields a conserved integral for radial solutions $u(t,r)$ iff we have 
$b=f(0)=0$ and $u_r|_{r=R} =0$, 
namely, a Neumann boundary condition on $u$ in the domain
(together with the regularity conditions \eqref{regularity}). 
Then the mass integral \eqref{mass} becomes 
an exponential function $M(t)=e^{kt} M_0$, 
where $M_0 = M(0)$ is the initial mass. 
Hence, the mass $M(t)$ either blows up when $k > 0$, 
or decays to zero when $k < 0$,
corresponding to when the term $f(u)=ku$ 
in the radial diffusion-reaction equation \eqref{nonlin_diffus_eqn_rad} 
acts as either a source or a sink.

The second conservation law \eqref{cons_law2_rad} 
involves a function $\phi(t,u)$ that satisfies the linear diffusion equation \eqref{lin_eqn_rad}. 
This indicates, similarly to the situation for the $1$-dimensional conservation law \eqref{cons_law1}, 
that the nonlinear radial diffusion-reaction equation \eqref{nonlin_diffus_eqn_rad} 
can be invertibly mapped into a linear diffusion equation by a hodograph transformation. 
In particular, 
this transformation is exactly the same as the one shown in Proposition~\ref{hodograph}.

Hence, 
from the classification of low-order radial conservation laws in Theorem~\ref{cons_laws_thm_rad}, 
combined with the general mapping theorem in \cite{AncBluWol}, 
we have the following result. 

\begin{prop}\label{hodograph_rad}
The radial nonlinear diffusion-reaction equation \eqref{nonlin_diffus_eqn_rad}
can be mapped into a linear diffusion equation by a point transformation iff
the diffusivity has the form $g(|u_r|)= -\kappa/|u_r|^2$
(with the source/sink $f(u)$ being arbitrary). 
The linear diffusion equation is given by 
$\tilde{u}_{\tilde{t}} = \kappa \tilde{u}_{\tilde{r}\tilde{r}} + f(\tilde{r})\tilde{u}_{\tilde{r}}$
where $(t,r,u) \rightarrow (\tilde{t},\tilde{r},\tilde{u})= (t,u,r^n/n)$
is the point transformation. 
\end{prop}

We next state the classification of Lie symmetries, 
excluding all hodograph cases. 

\begin{thm}\label{point_symm_thm_rad}
(i) 
The point symmetries admitted by 
the radial generalized $p$-Laplacian diffusion-reaction equation \eqref{nonlin_diffus_eqn_rad} 
for arbitrary $g(|u_r|)$
are generated by the transformations 
\eqref{point_symm1}--\eqref{point_symm5}. 
(ii) 
The radial generalized $p$-Laplacian diffusion-reaction equation \eqref{nonlin_diffus_eqn_rad}  
admits additional point symmetries only when 
$g(|u_r|)=-\kappa |u_r|^{2l/d}$ 
where $l$ is a non-zero integer and $d$ is a non-even number. 
These point symmetries (in non-hodograph cases)
are generated by the transformations 
\eqref{point_symm6}--\eqref{point_symm8} 
where $p=1+2l/d$. 
(iii) No contact symmetries other than prolongations of point symmetries 
are admitted by the radial generalized $p$-Laplacian diffusion-reaction equation \eqref{nonlin_diffus_eqn_rad}
(in non-hodograph cases). 
\end{thm}

By combining this result with Theorem~\ref{symm_groups}, 
we obtain a classification of point symmetry groups. 

\begin{thm}\label{symm_groups_rad}
The maximal point symmetry groups for the class of radial generalized $p$-Laplacian diffusion-reaction equation \eqref{nonlin_diffus_eqn_rad} 
(in all non-hodograph cases)
are generated by:
\begin{enumerate}[(1{-dimensional})]
\item\quad
\newline
arbitrary $g(|u_r|)$, $f(u)$:
\newline 
$\X_{1}$
\item\quad
\newline
arbitrary $g(|u_r|)$, $f(u)=b+ku$; $k\neq 0$: 
\newline
$\X_{1}$, $\X_{3}$, with commutator structure \eqref{2dim_1}
\newline
\newline
arbitrary $g(|u_r|)$, $f(u)=k/(a+u)$: 
\newline
$\X_{1}$, $\X_{4}$, with commutator structure \eqref{2dim_2}
\newline
\newline
$g(|u_r|)=-\kappa |u_r|^{p-1}$, 
$f(u)=k(a+u)^{p}+c(a+u)$; $c\neq 0$:
\newline
$\X_{1}$, $\X_{6}$, with commutator structure \eqref{2dim_3}
\newline
\newline
$g(|u_r|)=-\kappa |u_r|^{p-1}$, 
$f(u)=k(a+u)^{q}$; $q\neq -1,0$:
\newline
$\X_{1}$, $\X_{7}$, with commutator structure \eqref{2dim_4}
\newline
\newline
$g(|u_r|)=-\kappa |u_r|^{p-1}$, 
$f(u)=k\exp(qu)$; $q\neq 0$:
\newline
$\X_{1}$, $\X_{8}$, with commutator structure \eqref{2dim_5}
\item\quad
\newline
arbitrary $g(|u_r|)$, $f(u)=b$: 
\newline
$\X_{1}$, $\X_{3}|_{k=0}$, $\X_{5}$, with commutator structure \eqref{3dim_1}
\item\quad
\newline
$g(|u_r|)=-\kappa |u_r|^{p-1}$, 
$f(u)=b$:
\newline
$\X_{1}$, $\X_{3}|_{k=0}$, $\X_{5}$, $\X_{15}$, with commutator structure \eqref{4dim_1}
\newline
\newline
$g(|u_r|)=-\kappa |u_r|^{p-1}$, 
$f(u)=b+ku$: 
\newline
$\X_{1}$, $\X_{3}$, $\X_{6}|_{k=0,c=k,a=b/k}$, $\X_{7}|_{q=1,a=b/k}$, with commutator structure \eqref{4dim_3}
\end{enumerate}
where $p=1+2l/d$, $l$ is a non-zero integer and $d$ is a non-even number. 
\end{thm}

Finally, we remark that 
the radial generalized $p$-Laplacian diffusion-reaction equation \eqref{nonlin_diffus_eqn_rad} 
admits additional symmetries and conservation laws 
that hold on a subspace of the whole radial solution space. 
These are known as \emph{conditional} symmetries and conservation laws,
and they arise when any of the low-order local conservation laws and Lie symmetries 
shown in Theorems~\ref{cons_laws_thm} and~\ref{point_symm_thm}
for the $1$-dimensional diffusion-reaction equation \eqref{nonlin_diffus_eqn_1D}--\eqref{diffus_rel_1D} 
admit a restriction to solutions $u(t,r)$ such that $u_r=\pm|u_r|$ has a single sign. 
In particular, 
for such solutions, 
$g(|u_r|) = g(\pm u_r)= h(u_r)/u_r$ will be well-defined without the need to restrict $h(u_r)$, 
in which the $1$-dimensional equation coincides with the radial equation. 
(Note this sign condition will not hold in general for all solutions $u(t,r)$ of
the $1$-dimensional equation \eqref{nonlin_diffus_eqn_1D}--\eqref{diffus_rel_1D}.)

\section{Group-invariant solutions}\label{solns_sec}

Exact analytical solutions are of considerable interest as 
they can provide insight into asymptotic behavior, extinction or blow-up behavior, 
and they also give a means by which to test numerical solution methods. 

Each one-dimensional point symmetry group that is admitted by 
the radial generalized $p$-Laplacian diffusion-reaction equation \eqref{nonlin_diffus_eqn_rad} 
can be used to reduce the equation to obtain 
corresponding group-invariant solutions $u(t,r)$
\cite{Olv,BA,BCA}. 
The specific form for these solutions 
for a given symmetry with generator $\X$ 
is given by integration of the invariance condition $\hat{\X} |_{u(t,r)} = 0$, 
where $\hat{\X}$ is the symmetry generator
in characteristic form \eqref{Xchar_form}.

When the given symmetry $\X$ does not leave both variables $(t,r)$ invariant, 
the group-invariant solutions will have the form 
\begin{equation}\label{group-inv_soln}
u = \Phi(t,r,U(z))
\end{equation}
in terms of symmetry invariants 
\begin{equation}\label{invariants}
z=\zeta(t,r), 
\quad 
U=\Upsilon(t,r,u)
\end{equation}
given by 
\begin{equation}\label{invariance_condition}
\X z = \X U =0
\end{equation}
where $U(z)$ will satisfy a second-order nonlinear ODE obtained by 
reduction of the radial equation \eqref{nonlin_diffus_eqn_rad}. 
However, this ODE must still be solved to find $U(z)$, 
if explicit expressions for the group-invariant solutions are being sought. 
Not all second-order nonlinear ODEs can be solved explicitly, 
but there are many different integration methods that can be applied 
\cite{Olv,BA,AncFen}. 
One way to ensure that the ODE for $U(z)$ can be solved is 
if it inherits a two-dimensional symmetry group from the radial equation \eqref{nonlin_diffus_eqn_rad},
since then the inherited symmetries can be used \cite{Olv,BA} 
to reduce the ODE to a first-order separable form which can be directly integrated. 
The condition that the ODE inherits a two-dimensional symmetry group 
is equivalent to requiring that the starting one-dimensional symmetry group 
is a subgroup of a three-dimensional group of symmetries given by generators 
$\X$, $\Y_1$, $\Y_2$ with the commutator structure 
\begin{equation}\label{reduc_group_struct}
[\Y_1,\X]=C_1\X,
\quad
[\Y_2,\X]=C_2\X,
\quad
[\Y_2,\Y_1]=C_3\Y_1 
\end{equation}
where $C_1$, $C_2$, $C_3$ are constants. 
This structure describes a three-dimensional solvable Lie group \cite{BA},
with $\Y_1$ and $\Y_2$ generating a two-dimensional solvable Lie subgroup. 

Recall \cite{SagWal}, 
a Lie group $\G$ is solvable iff it has the structure 
$\G_{1}\subset \G_{2}\subset \cdots \subset \G$ 
in which $\G_{i}$ is a subgroup of dimension $i$ and a normal subgroup of $\G_{i+1}$, 
for $i=1,2,\ldots,\dim\G-1$. 
This structure has two equivalent formulations in the Lie algebra $\g$ of $\G$:
first, 
$\g_{1}\subset \g_{2}\subset \cdots \subset \g$ 
in which $\g_{i}$ is an $i$-dimensional normal subalgebra in $\g_{i+1}$, 
namely $\left[\g_{i+1},\g_{i}\right]\subseteq \g_{i}$,
$i=1,2,\ldots,\dim\g-1$;
second, 
$\g\supset \g^{(1)} \supset \g^{(2)} \supset \cdots \supset\g^{(\ell)}\supset \g^{(\ell+1)}=0$,
where $\g^{(i)} =[\g^{(i-1)},\g^{(i-1)}]$ is the $i$th derived subalgebra 
defined by induction, $i=1,2,\ldots,\ell\leq\dim\g$, 
with $\g^{(0)}=\g$. 
Note that the last derived subalgebra $\g^{(\ell)}$ is abelian,
and every derived subalgebra is an ideal in $\g$ 
as shown by a straightforward induction argument \cite{SagWal}. 
This second formulation will be the most useful one for the method of symmetry reduction
because it turns out to provide the required commutator structure \eqref{reduc_group_struct}, 
without any additional steps. 

We will now apply this symmetry reduction method to obtain exact analytic solutions 
for the radial generalized $p$-Laplacian diffusion-reaction equation \eqref{nonlin_diffus_eqn_rad} 
in $n>1$ dimensions, 
using the solvable symmetry groups of dimension three or higher 
admitted by this equation. 

From the classification of symmetry groups stated in Theorem~\ref{symm_groups_rad}, 
we begin by outlining how the commutator structures can be used to find 
all solvable symmetry subgroups of dimension at least three. 

For each given maximal symmetry algebra, $\g$, which is non-abelian, 
first, we find its derived subalgebras 
$\g^{(i)}$, for $i=1,2,\ldots,\dim\g$. 
We stop at $i=\dim\g$ because we know that $\g$ is solvable iff $\g^{(\dim\g)}$ is empty. 
Next, in the case when $\g^{(\dim\g)}$ is empty, 
if the derived subalgebra $\g^{(\ell)}$ is three-dimensional, 
it is abelian and therefore is trivially solvable. 
If instead $\g^{(\ell)}$ is two-dimensional, 
then the derived subalgebra $\g^{(\ell-1)}$ is a solvable (sub)algebra of dimension
at least three,
and finally, if $\g^{(\ell)}$ is one-dimensional, 
then either $\g^{(\ell-1)}$ or $\g^{(\ell-2)}$ is a solvable (sub)algebra of dimension
at least three. 

This procedure yields all possible three-dimensional solvable symmetry (sub)algebras
$\gsolv$  
and hence all possible corresponding solvable symmetry (sub)groups
$\Gsolv$. 
For each one, 
the generators in $\gsolv$ always can be arranged to have the commutator structure \eqref{reduc_group_struct}
where $\X$ is a symmetry generator in the abelian subalgebra $\g^{(\ell)}$,
and where $\g_{\X}=\gsolv/\spn(\X)$ is then a two-dimensional symmetry algebra 
that will be inherited by the second-order nonlinear ODE 
arising in the symmetry reduction given by $\X$. 
If $\g^{(\ell)}$ is one-dimensional,
then $\X$ is any generator of $\g^{(\ell)}$,  
and so $\g_{\X}=\gsolv/\g^{(\ell)}$ is unique. 
Next, if $\g^{(\ell)}$ is two-dimensional,
then two different cases can arise, 
depending on whether the adjoint action of $\Gsolv$ on $\g^{(\ell)}$ has orbits that are
one or two dimensional. 
In the one-dimensional case, 
$\X$ is any generator on either one of the two orbits in $\g^{(\ell)}$, 
whereas in the two-dimensional case, 
$\X$ is any generator in $\g^{(\ell)}$. 
Finally, if $\g^{(\ell)}$ is three-dimensional,
then again $\X$ is any generator in $\g^{(\ell)}$
since the orbits of the adjoint action of $\Gsolv$ on $\g^{(\ell)}$ 
are trivially three-dimensional. 

For a given solvable algebra of symmetries \eqref{reduc_group_struct}, 
whenever the set of allowed generators $\X$ is more than one-dimensional, 
any two of these symmetry generators that are related 
by the adjoint action of the full symmetry group $\G$ on $\g^{(\ell)}$ 
will produce group-invariant solutions that can be mapped into each other 
by the symmetry transformation on $(t,r,u)$ corresponding to this group action. 
Consequently, in this case it is sufficient to consider a subset of symmetry generators 
that belong to different equivalent classes under the adjoint action of $\G$. 
Such a subset is called an optimal set of symmetry generators
and can be straightforwardly determined by the methods in Refs.~\cite{Olv,WinPat}.
Note that when the set of generators $\X$ is only one-dimensional, 
then there is only a single equivalent class 
and hence $\X$ itself constitutes an optimal set. 
Also, as noted earlier, 
any generator $\X$ must have the property that 
it does not leave both variables $(t,r)$ invariant. 

To proceed, 
we now state a classification of solvable symmetry (sub)algebras $\gsolv$, 
and a corresponding optimal set of symmetry generators $\X$ 
along with the two-dimensional solvable symmetry subalgebras 
$\g_{\X}=\gsolv/\spn(\X) =\spn(\Y_1,\Y_2)$. 

\begin{prop}\label{solvable_symm_algebras}
For the radial generalized $p$-Laplacian diffusion-reaction equation \eqref{nonlin_diffus_eqn_rad} 
in $n>1$ dimensions, 
all solvable symmetry subalgebras \eqref{reduc_group_struct} are given by:
\begin{enumerate}[(a)]
\item 
arbitrary $g(|u_r|)$, 
$f(u)=b$:
\begin{gather}
\tilde\X_{1} = \partial_t, 
\quad
\tilde\X_{3}= \partial_u,
\quad
\tilde\X_{5} = 2t \partial_t + r \partial_r + (bt+u) \partial_u, 
\label{3-dim-X}
\\
\begin{aligned}
& 
\gsolv
 = \spn(\tilde\X_{1},\tilde\X_{3},\tilde\X_{5}) 
\supset
\g^{(1)} = \spn(\tilde\X_{1},\tilde\X_{3}),
\quad
\g^{(2)} = 0
\label{solvable-X1X3-X5}
\end{aligned}
\\
\X= \tilde\X_{1}+ b \tilde\X_{3}
\label{optimal-X1X3}
\\
\Y_1= \tilde\X_{3},
\quad
\Y_2=\tilde\X_{5}
\label{optimal-X1X3-symmalg}
\end{gather}
\item 
$g(|u_r|)=-\kappa |u_r|^{p-1}$, 
$f(u)=b+ku$, $p=1+2l/d$, $k\neq0$:
\begin{gather}
\begin{aligned}
& \tilde\X_{1} = \partial_t, 
\quad
\tilde\X_{3}= \exp(kt) \partial_u,
\quad
\tilde\X_{6} = \exp(k(1-p)t)( k^{-1} \partial_t + (k^{-1}b+u) \partial_u ), 
\\&
\tilde\X_{7} = (p-1)r \partial_r +(p+1)(k^{-1}b+u) \partial_u
\end{aligned}
\label{4-dim-X}
\\
\begin{aligned}
& \g = \spn(\tilde\X_{1},\tilde\X_{3},\tilde\X_{6},\tilde\X_{7}) 
\supset
\g^{(1)} = \spn(\tilde\X_{3},\tilde\X_{6}),
\quad
\g^{(2)} = 0
\label{solvable-X6X3-X1-X7}
\end{aligned}
\\
\X = k\tilde\X_{6}+\mu \tilde\X_{3},
\quad
\mu = \const
\label{optimal-X6X3}
\\
\Y_1 =\tilde\X_{3},
\quad
\Y_2=(p+1)\tilde\X_{1}+pk\tilde\X_{7}
\label{optimal-X6X3-symmalg}
\end{gather}
where $l$ is a non-zero integer and $d$ is a non-even number. 
\end{enumerate}
\end{prop}

We remark that once the resulting group-invariant solutions \eqref{group-inv_soln} have been found 
then in each case the full group $\G\supseteq\Gsolv$ of symmetry transformations 
can be applied to these solutions. 
However, since $\Gsolv$ is an ideal in $\G$, 
the group of symmetry transformations will map each family of group-invariant solutions 
into itself, and so no new solutions can be produced in this way.

\subsection{Reduction under combined time-translation and shift}

We first consider the $3$-dimensional solvable symmetry group \eqref{3-dim-X}--\eqref{solvable-X1X3-X5}. 
The optimal symmetry generator \eqref{optimal-X1X3} is explicitly given by 
\begin{equation}\label{reduc-X1X3}
\X=\tilde\X_{1} +b \tilde\X_{3}= \partial_t +b \partial_u,
\end{equation}
Its has the invariants \eqref{invariants}--\eqref{invariance_condition} given by 
\begin{equation}\label{inv-X1X3}
z=r,
\quad
U= u- b t
\end{equation}
from which we see that the form \eqref{group-inv_soln} for group-invariant solutions 
consists of 
\begin{equation}\label{group-inv-X1X3}
u=b t +U(r)
\end{equation}
Substitution of this expression into the $n$-dimensional radial diffusion-reaction equation \eqref{nonlin_diffus_eqn_rad}
yields the second-order nonlinear ODE 
\begin{equation}\label{ODE-X1X3}
(U'g(|U'|))' + (n-1)r^{-1} U'g(|U'|) =0 
\end{equation}
for $U(r)$, 
where $g$ is an arbitrary function, 
describing the diffusivity, 
and where $b$ is an arbitrary constant, 
describing a uniform, time-independent source/sink. 

From the other two symmetry generators \eqref{optimal-X1X3-symmalg}
in the solvable symmetry group, 
we have 
$\tilde\X_{3} r= 0$, $\tilde\X_{3} U= 1$, 
$\tilde\X_{5} r = r$, $\tilde\X_{5} U = U$. 
Thus, the ODE \eqref{ODE-X1X3} 
inherits a two-dimensional solvable symmetry algebra generated by 
\begin{equation}
\Y_1= \partial_U,
\quad
\Y_2 = r \partial_r + U\partial_U
\end{equation}
with the commutator structure $[\Y_2,\Y_1]=-\Y_1$. 
These symmetries allow the ODE \eqref{ODE-X1X3} to be integrated as follows.
First, we see that $\zeta=r$ and $V=U'$ are invariants of $\Y_1$,
and that $\Y_2\zeta = \zeta$ and $\pr^{(1)}\Y_2 V = 0$. 
In terms of these variables, the ODE becomes 
\begin{equation}
(Vg(|V|))' + (n-1)\zeta^{-1} Vg(|V|) =0
\end{equation}
which is a first-order ODE for $V(\zeta)$. 
It possesses $\pr^{(1)}\Y_2\big|_{(\zeta,V)}=\zeta\partial_\zeta$ as a symmetry,
and so it can be reduced to quadrature 
\begin{equation}
\int \Big( V^{-1} + \frac{g'(|V|)}{g(|V|)}\Big)\,dV =(1-n)\int \zeta^{-1}\,d\zeta +c_0,
\quad
c_0=\const
\end{equation}
Both integrals can be evaluated explicitly, yielding
$\zeta^{n-1}Vg(|V|) = e^{c_0}=c_1=\const$. 
We can write this solution in the explicit form 
\begin{equation}
V(\zeta) = h^{-1}(c_1\zeta^{1-n})
\end{equation}
where $h^{-1}$ denotes the inverse of the function 
\begin{equation}\label{h-X1X3}
h(V)= Vg(|V|)
\end{equation}

Next, after undoing the change of variables $V=U'$ and $\zeta=r$, 
we can integrate the resulting first-order ODE $U'= h^{-1}(c_1r^{1-n})$ 
to obtain the quadrature
\begin{equation}
U(r) = c_2 + \int h^{-1}(c_1 r^{1-n})\, dr,
\quad
c_1,c_2 =\const
\end{equation}
which yields the general solution of the second-order nonlinear ODE \eqref{ODE-X1X3}. 
Substituting this quadrature into the expression \eqref{group-inv-X1X3}, 
we finally get 
\begin{equation}\label{soln-X1X3}
u(t,r)=b t +c_2 + \int h^{-1}(c_1 r^{1-n})\, dr
\end{equation}
which gives all group-invariant solutions 
arising from the symmetry \eqref{reduc-X1X3},
where $h$ is the function \eqref{h-X1X3} and $h^{-1}$ is its inverse,
and where $c_1$, $c_2$ are arbitrary constants.

\subsection{Reduction under combined time-dependent dilation and shift}

We next consider the $4$-dimensional solvable symmetry group \eqref{4-dim-X}--\eqref{solvable-X6X3-X1-X7}. 
The optimal symmetry generator \eqref{optimal-X6X3} is explicitly given by 
\begin{equation}\label{reduc-X6X3}
\X=k\tilde\X_{6} +\mu \tilde\X_{3}
= e^{k(1-p)t} \partial_t + \big(e^{k(1-p)t}(b+ku) +\mu e^{kt}\big)\partial_u, 
\quad
p\neq1,
\quad
\mu=\const
\end{equation}
which has 
\begin{equation}\label{inv-X6X3}
z=r,
\quad
U= e^{-kt}(u +\tfrac{b}{k}) - \tfrac{\mu}{(p-1)k} e^{k(p-1)t}
\end{equation}
as invariants \eqref{invariants}--\eqref{invariance_condition}. 
We see that the form \eqref{group-inv_soln} for group-invariant solutions 
consists of 
\begin{equation}\label{group-inv-X6X3}
u=e^{kt}U(r)  + \tfrac{\mu}{(p-1)k} e^{kpt} -\tfrac{b}{k}
\end{equation}
Substitution of this expression into the $n$-dimensional radial diffusion-reaction equation \eqref{nonlin_diffus_eqn_rad}
yields the second-order nonlinear ODE 
\begin{equation}\label{ODE-X6X3}
\kappa ((U')^p)' + (n-1)r^{-1} (U')^p +\mu =0
\end{equation}
for $U(r)$. 

From the other two symmetry generators \eqref{optimal-X6X3-symmalg}
in the solvable symmetry group, 
we have 
$\tilde\X_{3} r= 0$, $\tilde\X_{3} U= 1$, 
$((p+1)\tilde\X_{1}+p\tilde\X_{7})r = p(p-1) k r$, 
$((p+1)\tilde\X_{1}+p\tilde\X_{7})U = p(p^2-1) k U$. 
Thus, the ODE \eqref{ODE-X6X3} 
inherits a two-dimensional solvable symmetry algebra generated by 
\begin{equation}
\Y_1= \partial_U,
\quad
\Y_2 = p r \partial_r + (p+1) U\partial_U
\end{equation}
(up to a scaling of $\Y_2$) 
with the commutator structure $[\Y_2,\Y_1]=k(1-p^2)\Y_1$. 
We use these symmetries to integrate the ODE \eqref{ODE-X6X3} as follows.
First, we see that $\zeta=r$ and $V=U'$ are invariants of $\Y_1$,
and that $\Y_2\zeta = p\zeta$ and $\pr\Y_2 V = V$. 
In terms of these variables, the ODE is given by 
\begin{equation}
\kappa (V^p)' + (n-1)\zeta^{-1} V^p +\mu =0
\end{equation}
for $V(\zeta)$. 
This is a first-order ODE which possesses 
$\pr^{(1)}\Y_2\big|_{(\zeta,V)}=p\zeta\partial_\zeta+V\partial_V$ as a symmetry. 
By changing variables from $V$ to the invariant $W=V^p/\zeta$, 
we see that this ODE becomes 
\begin{equation}
\zeta W' + nW +\mu/\kappa =0
\end{equation}
which has the quadrature 
\begin{equation}
\int \frac{dW}{nW +\mu/\kappa} =-\int \zeta^{-1}\,d\zeta +c_0,
\quad
c_0=\const
\end{equation}

Next, we evaluate the integrals, yielding
\begin{equation}
W(\zeta) = c_1\zeta^{-n} -\tfrac{\mu}{n\kappa} = V(\zeta)^p/\zeta, 
\quad
c_1=e^{c_0}=\const
\end{equation}
After undoing the change of variables $V=U'$ and $\zeta=r$, 
we can integrate the resulting first-order ODE 
$U'= ( (c_1r^{-n} -\mu/(n\kappa))r )^{1/p}$
to obtain the quadrature
\begin{equation}
U(r) = c_2 + \int ( c_1r^{1-n} -\tfrac{\mu}{n\kappa}r )^{1/p}\, dr,
\quad
c_1,c_2 =\const
\end{equation}
which yields the general solution of the second-order nonlinear ODE \eqref{ODE-X6X3}. 
Finally, substituting this quadrature into the expression \eqref{group-inv-X6X3}, 
we get 
\begin{equation}\label{soln-X6X3}
u(t,r)=
e^{kt}\Big( c_2 + \int ( c_1 r^{1-n} -\tfrac{\mu}{n\kappa}r )^{1/p}\, dr \Big) 
+ \tfrac{\mu}{(p-1)k} e^{kpt} -\tfrac{b}{k}
\end{equation}
which gives all group-invariant solutions 
arising from the symmetry \eqref{reduc-X6X3},
where $c_1$, $c_2$ are arbitrary constants.

\subsection{Exact solutions and their properties}

We will now look at the physical and analytical features of 
the two families of explicit group-invariant solutions \eqref{soln-X1X3} and \eqref{soln-X6X3} 
obtained for the radial generalized $p$-Laplacian diffusion-reaction equation \eqref{nonlin_diffus_eqn_rad}.

Consider the first solution family \eqref{soln-X1X3}, 
which we will write in the form 
\begin{equation}\label{soln-X1X3-fam}
u(t,r)=b t + r_1\int_{r_0/r_1}^{r/r_1}h^{-1}(z^{1-n})\, dz,
\quad
r_1,r_0=\const
\end{equation}
where $h^{-1}$ is the inverse of the function $h(u_r) = u_rg(|u_r|)$ 
given in terms of the arbitrary diffusivity function $g$,
and where $b$ is the constant source/sink. 
Here the radial diffusion-reaction equation \eqref{nonlin_diffus_eqn_rad} is given by 
\begin{equation}\label{nonlin_diffus_eqn_f=b}
u_t = (g(|u_r|) +|u_r|g'(|u_r|))u_{rr} + (n-1)r^{-1}g(|u_r|)u_r + b
\end{equation}
on the half-line $\Rnum^+$. 
These solutions \eqref{soln-X1X3-fam} describe a superposition of 
a uniform dynamical background 
$u_{\text{dyn}}(t)=bt$ 
and a static mass distribution 
\begin{equation}\label{U-soln-X1X3-fam}
u_{\text{static}}(r) = r_1\int_{r_0/r_1}^{r/r_1}h^{-1}(z^{1-n})\, dz
\end{equation}
which has two parameters $r_1$ and $r_0$. 
The dynamical part of the solution satisfies the simple evolution equation $u_t = b$,
while the static part of the solution satisfies the spatial equation 
\begin{equation}\label{nonlin_lapl_eqn_rad}
(g(|u_r|) +|u_r|g'(|u_r|))u_{rr} + (n-1)r^{-1}g(|u_r|)u_r =0, 
\quad 
r>0
\end{equation}
with no source/sink term. 
This spatial equation is a generalization of the $p$-Laplacian equation,
and so $u_{\text{static}}(r)$ represents an equilibrium mass distribution 
for the radial generalized $p$-Laplacian diffusion-reaction equation \eqref{nonlin_diffus_eqn_rad}. 
If we view both the equilibrium solution and the spatial equation 
as being radial reductions of the $n$-dimensional generalized $p$-Laplacian diffusion-reaction equation \eqref{nonlin_diffus_eqn}
without a source/sink term, 
then we find that this equilibrium solution has the meaning of a fundamental solution 
for the generalized Laplace equation 
\begin{equation}\label{nonlin_lapl_eqn}
\nabla \cdot (g(|\nabla u|) \nabla u) = \delta (x),
\quad
g'\neq 0,
\end{equation}
for $u(x)$ on $\Rnum^n$,
where $r=|x|$ 
and $\delta(x)$ is the Dirac delta distribution. 
To see this interpretation, 
we integrate equation \eqref{nonlin_lapl_eqn} 
over a ball $B(R)$ with radius $R>0$ in $\Rnum^n$. 
By using the divergence theorem 
and the property that the Dirac delta distribution has unit mass, 
we obtain 
$1= \int_{\partial B(R)} g(|\nabla u|) \nabla u\cdot\hat x\,dS 
= \vol(S^{n-1}) (r^{n-1}g(|u_r|) u_r)|_{r=R}$,
where $S^{n-1}$ denotes the unit-radius hypersphere in $\Rnum^n$. 
We now differentiate $u= u_{\text{static}}(r) = r_1\int_{r_0/r_1}^{r/r_1}h^{-1}(z^{1-n})\, dz$
to get $u_r = h^{-1}(r_1^{n-1}r^{1-n})$
whereby $r^{n-1}g(|u_r|) u_r = r^{n-1}h(u_r) = r_1^{n-1}$ is a constant. 
Hence, we must have $\vol(S^{n-1}) r_1^{n-1}=1$, 
which determines $r_1^{n-1}=1/\vol(S^{n-1})$. 
Then $u(x)= u_{\text{static}}(|x|) = \int_{r_0}^{|x|}h^{-1}(z^{1-n}/\vol(S^{n-1}))\, dz$
is the general solution of the generalized Laplace equation \eqref{nonlin_lapl_eqn}. 

Therefore, the family of group-invariant solutions \eqref{soln-X1X3-fam} 
can be viewed as satisfying the equations
\begin{equation}\label{split_nonlin_diffus_eqn_f=b}
u_t =b,
\quad
\nabla \cdot (g(|\nabla u|) \nabla u)  = c_0\delta (x),
\quad
g'\neq 0
\end{equation}
with $u=u(t,|x|)$ on $\Rnum^n$, 
where $\delta(x)$ is the Dirac delta distribution,
and where $c_0 = \vol(\partial B(r_1))$ is a constant. 

We can get some further insight into these solutions \eqref{soln-X1X3-fam} 
by using the mass conservation law \eqref{cons_law1_rad}. 
Again we will view $u=u(t,|x|)$ as an $n$-dimensional radial function on $\Rnum^n$. 
Integration of this conservation law over a ball $B(R)$ with radius $R>0$ in $\Rnum^n$ 
then yields a global continuity equation for the total mass
$M(t;B(R)) = \vol(S^{n-1})\int_{0}^{R} u(t,r) r^{n-1}dr$. 
In particular, we have 
\begin{equation}\label{f=b_mass_flux}
\frac{d}{dt} M(t; B(R)) 
= \vol(S^{n-1})( bR^n/n + r_1^{n-1} ) 
= \vol(B(R)) b + \vol(\partial B(r_1))
\end{equation} 
through use of the divergence theorem
and the relation $r^{n-1}g(|u_r|) u_r = r^{n-1}h(u_r) = r_1^{n-1}$. 
(In addition, we have used the well-known formula 
$\vol(B(r)) = \tfrac{1}{n}r^n\vol(S^{n-1})$.)
Note the flux in this mass continuity equation consists of 
a volume term $\vol(B(R)) b$ plus a hypersurface term $\vol(\partial B(r_1))$,
which arise respectively from the constant source/sink term $b$ 
and the Dirac delta term $\vol(\partial B(r_1))\delta(x)$ 
in equations \eqref{split_nonlin_diffus_eqn_f=b}. 
We can integrate the mass continuity equation \eqref{f=b_mass_flux}
to obtain an expression for the total mass 
\begin{equation}
M(t;B(R)) = (\vol(B(R)) b + \vol(\partial B(r_1)))t + M(0;B(R))
\end{equation} 
with 
\begin{equation}
M(0;B(R)) = \vol(S^{n-1})\int_{0}^{R} u_{\text{static}}(r) r^{n-1}dr
\end{equation} 
coming from the static part \eqref{U-soln-X1X3-fam} of the solutions \eqref{soln-X1X3-fam}. 

Now, consider the second solution family \eqref{soln-X6X3}. 
Here the radial diffusion-reaction equation \eqref{nonlin_diffus_eqn_rad} is given by 
\begin{equation}\label{diffus_eqn_f=b+ku}
u_t = -\kappa (p|u_r|^{p-1}u_{rr} + (n-1)r^{-1}|u_r|^{p-1}u_r)  + ku+ b 
\end{equation}
on the half-line $\Rnum^+$,
where $p-1$ is diffusivity power, 
$b$ is the constant source/sink,
and $k$ is the coefficient of the reaction term. 
Note, as indicated in Theorem~\ref{symm_groups_rad}, 
we can write $|u_r|^{p-1}= u_r^{p-1}$ when $p=1+2l/d\neq 1$, 
in terms of a non-zero integer $l$ and a non-even number $d$. 
We will write the solutions \eqref{soln-X6X3} in the form 
\begin{equation}\label{soln-X6X3-fam}
u(t,r)=
e^{kt}\int_{r_0}^{r} ( r_1^{n-1}z^{1-n}-\tfrac{\mu}{n\kappa}z )^{1/p}\, dz
+ \tfrac{\mu}{(p-1)k} e^{kpt} -\tfrac{b}{k},
\quad
r_1,r_0=\const
\end{equation}
where $\mu$, $r_1$, $r_0$ are free parameters. 
There are two main cases to consider, 
depending on the two parameters $\mu$ and $r_1$. 

First, we consider the case $\mu=0$. 
The solutions \eqref{soln-X6X3} are then explicitly given by 
\begin{equation}\label{soln-X6X3-fam-mu=0}
u(t,r)=e^{kt}U(r) -\tfrac{b}{k}
\end{equation}
with 
\begin{equation}\label{U-X6X3-fam-mu=0}
U(r) = \begin{cases}
r_1^{(n-1)/p} ( r^{1+(1-n)/p} - r_0^{1+(1-n)/p} ), 
& p\neq n-1
\\
\ln(r/r_0), 
& p = n-1
\end{cases}
\end{equation}
The constant term in $u(t,r)$ describes 
a static, uniform equilibrium solution $u_{\text{equil}} =-\tfrac{b}{k}$ 
of the radial diffusion-reaction equation \eqref{diffus_eqn_f=b+ku}. 
The non-constant term has a separable form with respect to $t$ and $r$. 
It can be viewed as a kind of Green's function 
$\tilde{u}(t,r)= e^{kt}U(r)$ 
for the radial diffusion-reaction equation 
\begin{equation}\label{nonlin_diffus_eqn_g=power}
\tilde{u}_t -k\tilde{u}+\kappa \nabla \cdot (|\nabla\tilde{u}|^{p-1}\nabla\tilde{u}) = \psi(t)\delta (x)
\end{equation}
on $\Rnum^n$, where $r=|x|$. 
In particular, we have $\psi(t)= r_1^{n-1} e^{kpt}$,
while by the same argument used for the previous solutions \eqref{soln-X1X3-fam}, 
we see that $U(r)$ satisfies the $p$-Laplacian equation on $\Rnum^n$, 
\begin{equation}\label{p_lapl_eqn}
\kappa \nabla \cdot (|\nabla U|^{p-1}\nabla U) = c_0\delta (x)
\end{equation}
where $c_0 = \vol(\partial B(r_1))$ is a constant,
and $B(R)$ denotes a ball of radius $R>0$ in $\Rnum^n$. 
However, the total mass of $\tilde{u}(t,r)= e^{kt}U(r)$ given by 
$M(t) = \vol(S^{n-1})\int_{0}^{\infty} e^{kt}U(r)\,dr$ is infinite,
since the integral $\int_{0}^{\infty} U(r) r^m\,dr$ does not converge 
for $U(r)$ given by expression \eqref{U-X6X3-fam-mu=0}. 
(We remark that a standard Green's function would have finite mass 
and would satisfy equation \eqref{nonlin_diffus_eqn_g=power} with $\psi(t)=\delta(t)$.)

Next, we consider the case $\mu\neq 0$ and $r_1=0$. 
The solutions \eqref{soln-X6X3} are now explicitly given by 
\begin{equation}\label{soln-X6X3-fam-mu<>0}
u(t,r)=e^{kt}U(r) +\Phi(t) -\tfrac{b}{k}
\end{equation}
with 
\begin{align}
\Phi(t) & = \tfrac{\mu}{(p-1)k} e^{kpt} 
\label{Phi-X6X3-fam-mu<>0-r1=0}
\\
U(r) & = \begin{cases}
\big(\tfrac{-\mu}{\kappa n}\big)^{1/p} \tfrac{p}{p+1} ( r^{1+1/p} - r_0^{1+1/p} ), 
& p\neq -1
\\
\tfrac{-\kappa n}{\mu} \ln(r/r_0), 
& p = -1
\end{cases}
\label{U-X6X3-fam-mu<>0-r1=0}
\end{align}
Since $r_1=0$, there is no longer any singularity in $u(t,r)$ that comes from 
a Dirac delta term at $r=0$ in the radial diffusion-reaction equation \eqref{diffus_eqn_f=b+ku}. 
One way to see this is from the mass conservation law \eqref{cons_law1_rad},
with $u=u(t,|x|)$ viewed as an $n$-dimensional radial function on $\Rnum^n$. 
If we integrate this conservation law over a ball $B(R)$ with radius $R>0$ in $\Rnum^n$,
and use the divergence theorem combined with the relation 
$r^{n-1}(u_r)^p = e^{pkt} r^{n-1}(U')^p = \tfrac{-\mu}{\kappa n} r^n$,
then we find that the mass flux is given by 
$-\vol(\partial B(R)) X|_{r=R} = \vol(B(R)) (be^{-kt} + \mu e^{(p-1)kt})$,
where $X= \kappa \exp(-kt) r^{n-1} ((u_r)^p + (b/n)r)$ 
is the radial flux in the mass conservation law \eqref{cons_law1_rad}. 
Since the mass flux consists only of volume terms, 
we conclude that the radial diffusion-reaction equation \eqref{diffus_eqn_f=b+ku} 
does not have a Dirac delta term at $r=0$. 

We can directly evaluate the total mass 
$M(t;B(R)) = \vol(S^{n-1})\int_{0}^{R} u(t,r) r^{n-1}dr$
by using expressions \eqref{soln-X6X3-fam-mu<>0}--\eqref{U-X6X3-fam-mu<>0-r1=0}. 
First we observe that the integral 
$\int_{0}^{R} U(r) r^{n-1}dr$ 
converges at $r=0$ if $p=-1$ or if $p\neq-1$ and $-1/p < n-1$. 
In these cases, we then have 
\begin{equation}
M(t;B(R)) = \vol(B(R)) \Phi(t) + e^{kt} M_0(B(R))
\end{equation}
where
\begin{equation}
M_0(B(R)) = \begin{cases}
\big(\tfrac{-\mu}{\kappa n}\big)^{1/p} \tfrac{p}{p+1} \vol(B(R))
\big( \tfrac{pn}{1+p(n+1)} R^{1+1/p} -r_0^{1+1/p} \big),
& p\neq -1
\\
\tfrac{-\kappa n}{\mu} \vol(B(R)) \big(\ln(R/r_0) -\tfrac{1}{n}\big)
& p = -1
\end{cases}
\end{equation}
However, the mass $M_0(B(R))$ is unbounded for $R\rightarrow\infty$ in both cases,
and so on $\Rnum^+$ 
the solutions \eqref{soln-X6X3-fam-mu<>0}--\eqref{U-X6X3-fam-mu<>0-r1=0}
describe the superposition of 
an infinite mass distribution $e^{kt}U(r)$ 
and a time-dependent, uniform background $\Phi(t)$. 

We will now show that $U(r)$ can be modified 
by the introduction of moving boundary $r=R(t)$ at which $u(t,R(t))+b/k=0$,
with the solutions $u(t,r)$ being given by 
\begin{equation}\label{soln_moving_boundary_X6X3-fam-mu<>0-r1=0}
u(t,r)=-\tfrac{b}{k} + 
\begin{cases}
e^{kt}U(r) +\Phi(t), & 
0\leq r\leq R(t)
\\
0, & 
r\geq R(t)
\end{cases}
\end{equation}
in terms of expressions \eqref{Phi-X6X3-fam-mu<>0-r1=0}--\eqref{U-X6X3-fam-mu<>0-r1=0}. 
These piecewise solutions will physically represent a moving interface front
in a constant background. 

Radial moving interfaces in $\Rnum^n$ can be generally described by \cite{Mei} 
a Stephan condition $\sigma(t) R(t)^{n-1}R'(t) = [X]_{R(t)}$,  
where $X$ is the radial mass flux 
and $[X]_{R(t)}$ is its jump at the interface $r=R(t)$. 
From the mass conservation law \eqref{cons_law1_rad},
we have 
$X= \kappa \exp(-kt) \big( r^{n-1} ((u_r)^p + (b/n)r) \big)|_{r=R(t)}$ 
which yields 
\begin{equation}
R'(t) = \tilde{\sigma}(t) R(t),
\quad
\tilde{\sigma}(t) = \tfrac{\mu}{n} \sigma(t) e^{(p-1)kt}
\end{equation}
We will determine $\sigma(t)$ from the boundary condition $u(t,R(t))+b/k=0$. 
Using expressions \eqref{soln-X6X3-fam-mu<>0}--\eqref{U-X6X3-fam-mu<>0-r1=0} with $r_0=0$, 
we find 
$U(R(t))=-e^{-kt}\Phi(t) = -\tfrac{\mu}{(p-1)k} e^{(p-1)kt}$
and hence 
\begin{equation}\label{moving_boundary_X6X3-fam-mu<>0-r1=0}
R(t) = R_0 e^{qkt}, 
\quad
q=\tfrac{p(p-1)}{p+1},
\quad
R_0 = \big(\tfrac{-\mu}{qk}\big)^{p/(p+1)}\big(\tfrac{\kappa n}{-\mu}\big)^{1/(p+1)}
\end{equation}
This gives $\tilde{\sigma}(t) = qk$ and so we obtain 
\begin{equation}
\sigma(t) = \tfrac{qnk}{\mu} e^{(1-p)kt}
\end{equation}
As a result, 
we have solutions \eqref{soln_moving_boundary_X6X3-fam-mu<>0-r1=0}
that describe a radial moving interface \eqref{moving_boundary_X6X3-fam-mu<>0-r1=0}
on a constant background. 
These solutions will be regular \eqref{regularity} at $r=0$ If $0<p < 1$,
in which they also will have finite mass 
\begin{equation}
\begin{aligned}
M(t) & 
= \vol(S^{n-1}) \int_{0}^{R(t)} (u(t,r)+b/k)r^{n-1}\,dr
\\&
= \vol(B(R(t))) \tfrac{\mu}{(p-1)k} \big(  1- \big(\tfrac{-\mu}{\kappa n}\big)^{1/p} 
\tfrac{pn}{1+p(n+1)} \big) e^{pkt} 
\end{aligned}
\end{equation}

Last, we consider the case $\mu\neq 0$ and $r_1\neq 0$. 
The solutions \eqref{soln-X6X3} again have the form \eqref{soln-X6X3-fam-mu<>0},
with $\Phi(t)$ being unchanged while $U(r)$ is given by the quadrature
\begin{equation}\label{U-X6X3-fam-mu<>0-r1<>0}
U(r) = \int_{r_0}^{r}( r_1^{n-1} z^{1-n} - \tfrac{\mu}{\kappa n} z )^{1/p}\,dz 
\end{equation}
Compared with the solutions in previous case, 
here $U(r)$ is less singular. 
An asymptotic expansion of the integral \eqref{U-X6X3-fam-mu<>0-r1<>0}
shows that $U(r)$ is regular on $\Rnum^+$ if $0<-p<1$
and that its total radial mass $M^U = \int_{0}^{\infty} U(r) r^{n-1}\,dr$ is finite
if $0<-p<1/(n+1)$. 
Moreover, similarly to the case $\mu=0$, 
we find that $U(r)$ satisfies a $p$-Laplacian equation on $\Rnum^n$ 
with a Dirac delta source term 
\begin{equation}\label{p_lapl_eqn_mu<>0}
\kappa \nabla \cdot (|\nabla U|^{p-1}\nabla U) +\mu = c_0\delta (x)
\end{equation}
where $c_0 = \vol(\partial B(r_1))$ is a constant,
and $B(R)$ denotes a ball of radius $R>0$ in $\Rnum^n$. 
Thus, $U(r)$ is the fundamental solution of this one-parameter ($\mu\neq0$) family of 
$p$-Laplacian equations \eqref{p_lapl_eqn_mu<>0}. 

Consequently, 
if we subtract the constant background term from $u(t,r)$, 
we obtain solutions $\tilde{u}(t,r)= e^{kt}U(r) +\Phi(t)$ 
to a radial diffusion-reaction equation \eqref{nonlin_diffus_eqn_g=power}
on $\Rnum^n$ with a source term $\psi(t)\delta(x)$, 
where $r=|x|$ and $\psi(t)= r_1^{n-1} e^{kpt}$. 
These solutions will be regular and have finite mass 
$M(t) = \vol(S^{n-1}) M^U e^{kt}$ 
when the diffusivity power satisfies $0<-p<1/(n+1)$. 

For positive diffusivity powers, 
$\tilde{u}(t,r)= e^{kt}U(r) +\Phi(t)$ will be a singular solution,
as $U(r)$ will be unbounded for large $r$ when $p>0$
and also will be singular at $r=0$ if $p\leq m$. 
To obtain a solution with better regularity, 
we can seek a piecewise solution in which $U(r)$ has a cutoff at some radius $r=R$:
\begin{equation}\label{cutoff-X6X3-fam-mu<>0-r1<>0}
U(r)=U_0=\const \text{ for } r\geq R, 
\quad
U'(R)=0,
\quad
U''(R)=0
\end{equation}
These conditions will ensure that $U(r)$ is still a classical solution of 
the $p$-Laplacian equation \eqref{p_lapl_eqn_mu<>0} on $\Rnum^n$ for $r>0$. 
From expression \eqref{U-X6X3-fam-mu<>0-r1<>0}, 
we have 
$U'(r) = ( r_1^{n-1} z^{1-n} - \tfrac{\mu}{\kappa n} z )^{1/p}$
and 
$U''(r) = p^{-1}(1-n)( r_1^{n-1} z^{1-n} - \tfrac{\mu}{\kappa n} z )^{(1-p)/p} ( r_1^{n-1} z^{-n} - \tfrac{\mu}{\kappa n(1-n)} )$,
where we are assuming $p>0$. 
Imposing $U'(R)=0$, we get 
\begin{equation}\label{R-X6X3-fam-mu<>0-r1<>0}
R = \big( \tfrac{\kappa n}{\mu}r_1^{n-1} \big)^{1/n}
\end{equation}
Then we find 
$U''(R) = \tfrac{\mu}{p\kappa}(0)^{(1-p)/p}$, 
which will vanish if $1>p$. 
Thus, 
under the condition $0<p<1$ on the diffusivity power, 
we obtain a piecewise classical solution for $U(r)$, 
with a cutoff \eqref{cutoff-X6X3-fam-mu<>0-r1<>0}--\eqref{R-X6X3-fam-mu<>0-r1<>0}. 
(Note we will also need $\kappa/\mu>0$ when $n$ is even.)
This solution will have a finite radial mass 
$M^U = \int_{0}^{R} U(r) r^{n-1}\,dr$ 
if $U_0=U(R)=0$ and if $p>(n-1)/(n+1)$.  

As a result, for positive diffusivity powers $0<p<1$,
we have classical piecewise solutions of the form 
\begin{align}
u(t,r) & =e^{kt}U(r) +\Phi(t) -\tfrac{b}{k}
\label{soln-X6X3-fam-mu<>0-r1<>0}
\\
\Phi(t) & = \tfrac{\kappa n}{k(p-1)} r_1^{n-1}R^{-n} e^{kpt} 
\label{Phi-X6X3-fam-mu<>0-r1<>0}
\\
U(r) & = \begin{cases}
r_1^{(n-1)/p}  \int_{r_0}^{r} (z^{1-n} - R^{-n}z )^{1/p}\,dz, 
& 0\leq r\leq R
\\
r_1^{(n-1)/p}  \int_{r_0}^{R} (z^{1-n} - R^{-n}z )^{1/p}\,dz, 
& r\geq R
\end{cases}
\label{U-cutoff-X6X3-fam-mu<>0-r1<>0}
\end{align}
where the cutoff radius $R>0$ is arbitrary. 
If we put $r_0=R$, 
then these solutions describe the superposition of 
a mass distribution $e^{kt}U(r)$ with finite support 
and a time-dependent, uniform background $\Phi(t) -b/k$ 
on $\Rnum^+$. 
The mass distribution will have a finite total mass 
$e^{kt}M^U = \vol(S^{n-1})\int_{0}^{R} U(r)\,dr$
when the diffusivity power satisfies $1>p>(n-1)/(n+1)$.  

To conclude our discussion of the group-invariant solutions 
\eqref{soln-X6X3-fam-mu<>0-r1<>0}--\eqref{U-cutoff-X6X3-fam-mu<>0-r1<>0}
for $0<p<1$ 
and 
\eqref{soln-X6X3-fam-mu<>0-r1<>0}, \eqref{U-X6X3-fam-mu<>0-r1<>0}, 
\eqref{Phi-X6X3-fam-mu<>0-r1=0}
for $0<-p<1$, 
we remark that the integrals for $U(r)$ can be evaluated in terms of 
elementary functions when $n=2$ and hypergeometric functions when $n\neq2$:
\begin{equation}
U(r) = \begin{cases}
\begin{aligned}
& \tfrac{p(n-2)}{p+1} \big(\tfrac{-\mu}{n\kappa}\big)^{1/p} \Big( 
r^{1+1/p} F(\tfrac{-1}{p},\tfrac{p(n-2)}{p+n-1}; \tfrac{p(n-1)+1}{p(n-2)}; \tfrac{n\kappa}{\mu}r_1^{1-n} r^{n-2})
\\&\qquad
- r_0^{1+1/p} F(\tfrac{-1}{p},\tfrac{p(n-2)}{p+n-1}; \tfrac{p(n-1)+1}{p(n-2)}; \tfrac{n\kappa}{\mu}r_1^{1-n} r_0^{n-2}) \Big), 
\end{aligned}
& n\neq 2
\\
\tfrac{p}{p+1} r_1^{-1/p} (1- \tfrac{1}{2}\mu r_1\kappa^{-1})^{1/p} ( r^{1+1/p} - r_0^{1+1/p} ),
& n =2, p\neq -1
\\
r_1\kappa (\kappa- \tfrac{1}{2}\mu r_1)^{-1} \ln(r/r_0),
& n =2, p= -1
\end{cases}
\end{equation}

\section{Concluding Remarks}\label{remarks_sec}

The generalized $p$-Laplacian evolution equations studied in the present work 
describe an interesting class of radial nonlinear diffusion-reaction equations 
in $n\geq 1$ dimensions with many physical applications. 
The analysis of the initial-value problem for these equations is also interesting as 
solutions are known to exhibit several kinds of behaviour, 
such as finite-time blow up or finite-time extinction and moving interfaces. 

In future work, 
we plan to use of all the admitted Lie symmetries systematically 
to look for explicit solutions that display these different features. 
We also plan to extend our complete classification of 
low-order conservation laws and Lie symmetries 
to the non-radial case in $n>1$ dimensions.

\section*{Appendix}

Here we summarize the main steps that have been used in 
solving the determining systems \eqref{mult_det_sys}--\eqref{symm_det_sys}
for multipliers and symmetries. 
Because in the literature there are many papers that miss solutions and give wrong results for classification problems, 
this discussion may be helpful in a wider context. 

(1) 
Split the determining system by Maple ``rifsimp'', 
which yields a tree of all solution cases consisting of ODEs for $h$ and $f$, 
and a system of PDEs for $Q$ or $(\eta,\tau,\xi)$. 

(2) 
Solve the resulting ODEs and PDE system in each case 
by Maple ``dsolve'' and ``pdsolve''. 
(In particular, for every case, 
check that all solutions representing special branches of the general solution
of the ODEs and PDEs are obtained.)
Each solution case will consist of a linear set of $\{Q\}$ or $\{(\eta,\tau,\xi)\}$, 
along with a corresponding set of $\{h,f,m\}$. 

(3) 
Merge the solution sets in different cases by total inclusion. 
The resulting solution cases may still have some (partial) overlap. 

(4) 
Sort the solution cases into a case tree arranged by, firstly, the dimension of 
the linear sets of $\{Q\}$ or $\{(\eta,\tau,\xi)\}$,
and secondly, by inclusion or generality of $h,f,m$. 
(If there is a case where $h,f,m$ are arbitrary, then this is the top of the case tree.
Otherwise, the tree starts from the case, or cases, of greatest generality.)

(5) 
The case tree provides a full classification of all linear spaces of 
conservation laws or point symmetry algebras, 
arranged by dimensionality. 
An equivalent classification can be obtained by rearranging the solution cases 
strictly by generality of $h,f,m$. 
Any overlapping cases in the tree can be separated if a set of mutually exclusive solution cases is being sought. 

A more concise presentation of the classification can be obtained by listing 
a basis for all of the linear spaces of $\{Q\}$ or $\{(\eta,\tau,\xi)\}$ 
in the case tree. 
These linear spaces should be maximal (namely, they cannot be enlarged) in each case. 

(6) 
Choose a basis for the maximal linear space(s) of smallest dimension 
at the top of the case tree. 
For each subsequent case in the tree, 
choose a basis for the maximal linear space 
by extending the basis inherited from previous overlapping case(s). 
Wherever the extension is more than one-dimensional, 
a choice for the extended basis is most usefully made by considering 
the physical/geometrical meaning of the conservation laws or the point symmetries
represented by the basis,
as well as their properties under any scaling (or other important) transformations. 

(7)
Any cases that involve a free function should be handled separately, 
since they will correspond to the existence of 
a linearizing (or partially linearizing) transformation, 
depending on the number of variables in the free function(s). 

(8) 
Arrange the full basis into a table ordered by extension followed by generality of $h,f,m$ 
(or other important properties). 
This provides a complete set of generators for all of 
the admitted maximal Lie algebras of symmetries
or the admitted maximal vector spaces of conservation laws. 

We remark that these steps are applicable in general to classification problems 
that do not involve the question of equivalence transformations. 
For equivalence classification problems, additional preliminary steps are necessary,
which are fully discussed in Ref.\cite{PopIva}.

\end{document}